\documentclass[useAMS,usenatbib]{emulateapj}

\usepackage{lscape}

\newcommand{\Change}[1]{{#1}}

\newcommand{\ud}{\mathrm{d}}
\newcommand\Msun{\hbox{$M_{\odot}$}}
\newcommand\Lsun{\hbox{$L_{\odot}$}}
\newcommand\kms{\mbox{km~s$^{-1}$}}

\newcommand\DmaxVal{1.63}
\newcommand\DmaxErr{0.03}
\newcommand\VmaxVal{399}
\newcommand\VmaxErr{8}

\newcommand\DstdVal{0.89}
\newcommand\VstdVal{350}

\newcommand\DensityContrastMax{12}
\newcommand\DensityContrastStd{80}

\newcommand\NumGroupsMax{1538}

\newcommand\Dmax{\DmaxVal~Mpc}
\newcommand\Vmax{\VmaxVal~\kms}
\newcommand\Dstd{\DstdVal~Mpc}
\newcommand\Vstd{\VstdVal~\kms}

\newcommand\DCMax{{$\delta \rho / \rho$ = \DensityContrastMax}} 
\newcommand\DCStd{{$\delta \rho / \rho$ = \DensityContrastStd}}

\newcommand\hvalue{$h = 0.73$}
\newcommand\HAssumption{We assume {\hvalue}.}

\newcommand\ParamMax{{($D_0$, $V_0$) = (\Dmax, \Vmax)}}
\newcommand\ParamStd{{($D_0$, $V_0$) = (\Dstd, \Vstd)}}

\slugcomment{Accepted for publication in ApJ. Preprint 2006 Oct 30.}

\shorttitle{Groups of galaxies in 2MRS}
\shortauthors{A. C. Crook et al.}

\begin{document}

\title{Groups of galaxies in the Two Micron All-Sky Redshift Survey}

\author{Aidan C. Crook}
\affil{Kavli Institute for Astrophysics and Space Research, Massachusetts Institute of Technology, Cambridge, MA 02139}
\email{acc@space.mit.edu}

\author{John P. Huchra\altaffilmark{1}, Nathalie Martimbeau and Karen L. Masters}
\affil{Harvard-Smithsonian Center for Astrophysics, Cambridge, MA 02138}

\author{Tom Jarrett}
\affil{Infrared Processing and Analysis Center, California Institute of Technology, Pasadena, CA 91125}

\author{Lucas M. Macri}
\affil{National Optical Astronomy Observatories, Tucson, AZ 85719}

\altaffiltext{1}{Visiting Astronomer, Cerro Tololo Inter-American Observatory.
CTIO is operated by AURA, Inc.\ under contract to the National Science
Foundation.}

\begin{abstract}
We present the results of applying a percolation algorithm to the initial release of the Two Micron All-Sky Survey Extended Source Catalog, using subsequently measured redshifts for almost all of the galaxies with $K < 11.25$~mag.
This group catalog is based on the first near-IR all-sky flux-limited survey that is complete to $|b| = 5^{\circ}$.
We explore the dependence of the clustering on the length and velocity scales involved.
The paper describes a group catalog, complete to a limiting redshift of $10^4$~\kms, created by maximizing the number of groups containing 3 or more members.
A second catalog is also presented, created by requiring a minimum density contrast of $\delta \rho / \rho \ge 80$ to identify groups.
We identify known nearby clusters in the catalogs and contrast the groups identified in the two catalogs.
We examine and compare the properties of the determined groups and verify that the results are consistent with the UZC-SSRS2 and northern CfA redshift survey group catalogs.
The all-sky nature of the catalog will allow the development of a flow-field model based on the density field inferred from the estimated cluster masses.
\end{abstract}


\section{Introduction}

The Two Micron All-Sky Survey (2MASS) began in the early 90s with the purpose of mapping the Milky Way and nearby Universe. Previous all-sky surveys suffered from a variety of selection effects, many of them inconsistent across the sky. Observations at optical wavelengths suffer from severe extinction at low galactic latitudes, motivating work using surveys conducted in the infrared. Many IRAS-selected galaxy samples have been investigated as tracers of the galaxy density field, e.g. \citet{Strauss:1992a,Fisher:1995}, based on the 1.9 Jy and 1.2 Jy samples, respectively, and \citet{Branchini:1999} based on the PSCz catalog \citep{Saunders:2000}, however these samples are based on fluxes in the far-infrared and miss many early-type galaxies, thus underestimate the total galaxy number density. Even though the IRAS-selected samples are not biased by extinction, they still suffer from confusion in high-density regions.

Since galaxies' spectra peak at $\sim$1.6 $\mu$m, a survey in the near-infrared is optimized for their detection at the flux limit of the survey. By sampling uniformly over the entire celestial sphere in the J (1.25 $\mu$m), H (1.65 $\mu$m) and K (2.16 $\mu$m) bands, 2MASS has been designed to maximize the number of galaxies detected at a specified flux limit, producing the most complete all-sky survey performed to date \citep[2MASS,][]{Skrutskie:2006}.

Two catalogs, complete to $K\sim 13.5$~mag, were released in early 2003 (see the 2MASS explanatory supplement \citealt{Cutri:2003}, and \citealt{Jarrett:2000}): a point-source catalog with 470,992,970 entries and an extended source catalog (XSC) with 1,647,559 objects classified as galaxies. Although designed for completeness down to low galactic latitudes, the 2MASS XSC still suffers from confusion near the galactic plane. The 2MASS Redshift Survey \citep[hereafter 2MRS,][]{Huchra:2005,Huchra:2005b} uses the XSC as its input master list and aims to produce an all-sky, (extinction-corrected) flux-limited redshift catalog that will eventually be complete to $K=13.0$~mag above $|b|=5^\circ$. 
% In Prep reference
The 2MRS is currently 99.9\% complete to $K=11.25$~mag and $|b|>5^\circ$.\footnote{See 2MRS data release, Huchra et al. (2006), \textit{In preparation}.} 

In this paper, we create a redshift-limited catalog of groups, uniformly sampled from the entire sky. By assuming the identified groups are virialized systems, we are able to provide estimates of the group masses, avoiding the necessity to assume an intrinsic mass-to-light ratio.
The local Universe is sufficiently inhomogeneous at the scales in question that the dynamics due to our interactions with nearby groups are non-negligible. Due to the nature of the all-sky group catalog presented here, we will now be able to estimate the local density field due to baryonic matter in the local Universe. A flow-field model produced from this catalog can be used in conjunction with observations in order to answer the question of whether baryonic matter is a genuine tracer of dark matter.

The creation of group catalogs is not a new concept, however the methods employed in developing these catalogs have evolved with the enhancements in instrumentation. Early group catalogs were based on limited or subjective data \citep[e.g][]{deVaucouleurs:1975}, associating members based on similarity in apparent magnitude, positional coincidence and (if available) redshift. \citet{Turner:1976} proposed a method that identifies regions in which the surface number density on the sky is enhanced, creating group catalogs from two-dimensional data. This technique suffers because the typical angular separation of galaxies in a group will vary with distance, thus nearby groups with large angular radii will not be identified. Furthermore, when applied to flux-limited surveys, this method will identify different groups for different limiting fluxes. 

More recently, the use of objective algorithms to identify groups based on both their position on the sky and in redshift space has become widely accepted \citep[e.g.][]{Huchra:1982,Geller:1983,Ramella:1997,Diaferio:1999,Giuricin:2000,Ramella:2002}, using methods designed to find the same groups regardless of the limiting magnitude of the sample. The applicability of a particular group-finding algorithm depends on the properties of the sample in question. For example, \citet{Marinoni:2002} show that the Vornoi-Delauney method successfully reproduces the distribution of groups in velocity dispersion in a mock sample based on the Deep Extragalactic Evolutionary Probe (DEEP2) Redshift Survey \citep{Davis:2003}; this method is adapted by \citet{Gerke:2005} for application to the DEEP2 sample. The SDSS team developed an algorithm \citep[C4,][]{Miller:2005} that searches for groups in three space-dimensions as well as four photometric colors. \citet{Kochanek:2003} used a matched filter algorithm to study clusters in 2MRS at the 89\% completeness level. \Change{\citet{Yang:2005} have developed a halo-based group-finder and successfully applied it to the 2dFGRS sample \citep{Merchan:2002,Eke:2004}. The same technique has been applied to the SDSS by \citet{Weinmann:2006}.}

In this paper we apply a variable linking-length percolation (also commonly referred to as a \textit{friends-of-friends}) algorithm \citep[][hereafter HG82]{Huchra:1982} to determine the groups present in 2MRS. The velocity dispersion within the groups will allow estimates of the virial masses of the groups, thus providing a method to trace the density field associated with luminous matter in the local Universe.

We begin with an outline of the group-identification algorithm in \S\ref{Section:GroupAlgorithm} below. We discuss the modifications made to the data sample prior to the application of the algorithm in \S\ref{Section:DataModifications}. \S\ref{Section:ApplyAlgorithm} presents a discussion on the choice of parameters used in the group-identification algorithm. The group catalogs and their properties are discussed in \S\ref{Section:Groups}, and we summarize our conclusions and discuss the potential applications of the catalogs in \S\ref{Section:Summary}.

\section{Group-identification algorithm} \label{Section:GroupAlgorithm}

We use the algorithm described in HG82 to identify groups of galaxies in the $K<11.25$~mag version of the 2MRS catalog. The procedure is outlined briefly below.

We compare each galaxy in the catalog with its neighboring galaxies; for each pair of galaxies a linking length $D_L(V_\mathrm{avg})$ is computed that depends on the average redshift of the galaxies, $V_\mathrm{avg}$. Given two galaxies with an angular separation, $\theta$, we ask whether their projected separation, \mbox{$D_{12} = \sin (\theta/2) V_\mathrm{avg} / H_0$}, is less than $D_L(V_\mathrm{avg})$. If this is true, and the difference in redshift, \mbox{$V_{12} = |V_1-V_2|$}, is less than some linking velocity, $V_L$, then we identify both galaxies with the same group. $D_L$ is defined through equation (\ref{Eq:DScaling}) below.
\begin{equation} \label{Eq:DScaling}
D_L = D_0 \left[\frac{ \int_{-\infty}^{M_{12}(V_\mathrm{avg})} \Phi(M) \ud M }{ \int_{-\infty}^{M_\mathrm{lim}} \Phi(M) \ud M } \right]^{-1/3}
\end{equation}
where
\begin{displaymath}
M_{12}(V_\mathrm{avg}) = m_\mathrm{lim}-25-5 \log (V_\mathrm{avg}/H_0)
\end{displaymath}
Here, $\Phi(M)$ represents the differential galaxy luminosity function for the sample and $D_0$ the projected separation (in Mpc) at some chosen fiducial redshift $V_F$. $M_\mathrm{lim}=M_{12}(V_F)$ is a constant for a given $V_F$, and $m_\mathrm{lim}$ is the apparent-magnitude limit of the sample.

This scaling of the linking length compensates for the bias that would otherwise be introduced due to the variation in sampling of the luminosity function with redshift. 
There is much debate on how and whether or not to scale $V_L$ \citep[e.g., HG82,][]{Nolthenius:1987,Frederic:1995a,Frederic:1995b}.
If one assumes uniform density, simple scaling arguments show that the velocity is simply proportional to the radius, suggesting that $V_L$ should be scaled in the same manner as $D_L$. Such a scaling would include unwanted interlopers at large values of $V_L$ and thus introduce an unwanted correlation between velocity dispersion and redshift.
The density profiles of galaxy clusters, however, are usually better described by the isothermal-sphere approximation; in this case the velocity dispersion is independent of the size of the cluster. It follows therefore that by setting $V_L$ to a reasonable fixed value, we will minimize the number of interlopers, but not bias the algorithm against finding distant groups. Hereafter we set
\begin{equation} \label{Eq:VScaling}
V_L = V_0
\end{equation}

The choice of $D_0$ determines the minimum density contrast of identified groups, which can be estimated using equation (\ref{Eq:DensityContrast}) below (HG82).
\begin{equation} \label{Eq:DensityContrast}
\frac{\delta \rho}{\rho} = \frac{3}{4 \pi D_0^3} \left[ \int_{-\infty}^{M_\mathrm{lim}} \Phi(M) {\ud}M \right]^{-1} - 1
\end{equation}

\section{The sample} \label{Section:DataModifications}

The first available sample of the 2MRS galaxy catalog contains positions, redshifts and magnitudes for 23090 galaxies selected from the XSC. The targets were by selected by introducing a cut on the corrected magnitudes of objects in the XSC of $K < 11.25$~mag (the apparent magnitudes had previously been corrected for extinction using the dust maps of \citealt{Schlegel:1998}). This catalog is complete, bar 40 galaxies, for galactic latitudes \mbox{$|b| > 10^\circ$} between galactic longitudes 330$^\circ$ and 30$^\circ$, and \mbox{$|b| > 5^\circ$} for other longitudes.

Below, we discuss a simple flow-field model applied to provide improved estimates of the distances to the galaxies (see \S\ref{SubSection:Flowfield}). In \S\ref{SubSection:PopulatePlane} we discuss the method used to populate the galactic plane with random galaxies to prevent any artifacts arising from the significantly reduced observed number density of galaxies behind the plane. We briefly discuss the assumed form of the luminosity function of the sample in \S\ref{SubSection:LumFunc} and, in \S\ref{SubSection:VelocityLimit}, consider the completeness of the sample in redshift-space.

\subsection{Distance estimates} \label{SubSection:Flowfield}
Locally redshifts do not provide a reliable indication of distance because of distortions to the local velocity field due to infall onto concentrations of mass. Although the clustering algorithm is independent of the observer's frame of reference, it is essential to have reasonable estimates of the distances to the galaxies in order to compute the linking parameters, $D_{12}$ and $D_L$, as well to accurately estimate the luminosities of the galaxies.

We apply the basic flow-field model described in \citet{Mould:2000} to account for the local distortions to the velocity field.
This prescription first corrects the reference frame to the LG frame (\citealt*{Yahil:1977}, corroborated by the more recent work of \citealt*{Courteau:1999}), then adjusts the redshift-inferred distances of galaxies near Virgo, Shapley and the GA region as follows: All galaxies within 12$^\circ$ of the center of Virgo with heliocentric redshifts less than 2500~\kms~are placed at the redshift of Virgo (plus a random velocity, drawn from a gaussian distribution with a standard deviation of 20~\kms, to avoid artifacts in the group properties occurring from galaxies with identical redshifts). All galaxies within 10$^\circ$ and 2000~\kms~of the GA are placed at the redshift of the GA (plus scatter) and all galaxies within 12$^\circ$ and 3000~\kms~of Shapley are placed at the redshift of Shapley (plus scatter). The corrected velocities are then used in place of the heliocentric velocities when computing distances only.
To infer the distances we assume Hubble's law is valid to the completeness limit of the 
2MRS catalog,\footnote{At the limiting redshifts of the galaxies analyzed, the difference between distances computed using a $\Lambda$CDM cosmology and simply assuming Hubble's law is less than 5\%.}
using a Hubble constant $H_0 = 100h$ {\kms}Mpc$^{-1}$, where we assume \hvalue~when a specific value is required. This value is chosen based on the three-year WMAP results \citep{Spergel:2006}, $h=0.73 \pm 0.03$. In the very local universe (i.e. where corrected distances are less than $3h^{-1}$ Mpc) we give galaxies an indicative distance of $3h^{-1}$ Mpc. The velocities used in computation of $V_{12}$, etc. are the heliocentric velocities reported in the 2MRS catalog.

\subsection{Filling in the galactic plane} \label{SubSection:PopulatePlane}
The 2MRS catalog is currently incomplete near the galactic plane (\mbox{$|b| < 10^\circ$} between galactic longitudes 330$^\circ$ and 30$^\circ$, and \mbox{$|b| < 5^\circ$} for other longitudes). With a significantly reduced number density of galaxies observed behind the galactic plane, a structure that spans the plane will not be identified by the clustering algorithm. Similarly, structures that are visible in part above or below the plane may not be identified as groups, and, even if they are, a bias will be introduced in the number density of groups with centers just above or below the plane. Any flow-field model derived from such a group catalog will suffer from these biasing effects; we therefore attempt to minimize these effects by randomly populating the sample to enhance the galaxy number density behind the galactic plane to reflect that observed above and below it.

\begin{figure*}
\begin{center}
\includegraphics[width=0.9\textwidth]{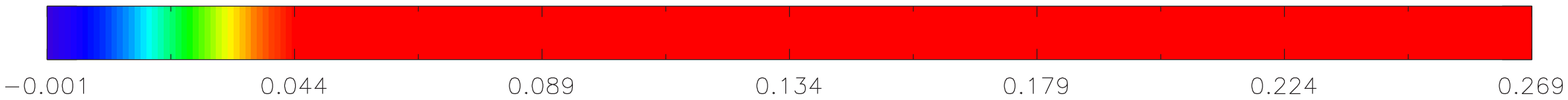} \\
\vspace{5mm}
\includegraphics[width=\textwidth]{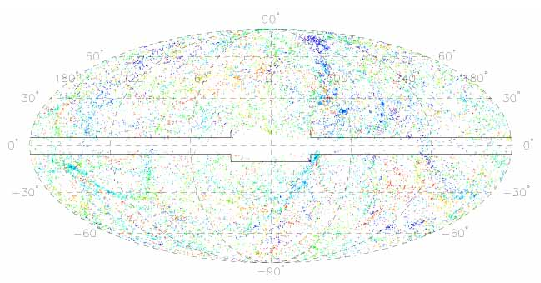} \\
\vspace{5mm}
\includegraphics[width=\textwidth]{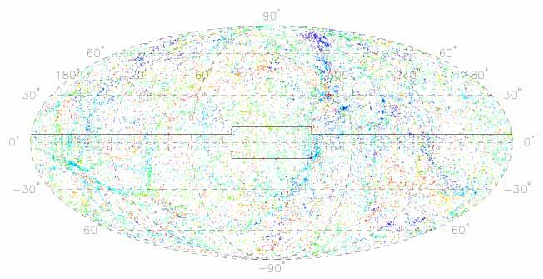}
\end{center}
\caption{\label{Fig:PopulatePlane}Galaxies in 2MRS catalog shown in a Mollweide projection in galactic coordinates. The top panel shows the catalog before the plane was populated. The bottom panel shows the catalog including the addition of the randomly-generated galaxies. The solid line indicates the region that was populated. (The color in the electronic edition indicates the measured redshift of the galaxy.)
}
\end{figure*}

We follow a method similar to that of \citet{Yahil:1991}; this method produces similar results to the more involved 
% In Prep Reference
Wiener reconstruction \footnote{See \citet{Erdogdu:2006b}, and Rassat et al. (2006), \textit{in preparation} for discussions on the Wiener reconstruction of the 2MRS sample.}
discussed in \citet{Lahav:1994}. We first divide the catalog into bins spanning $10^\circ$ in galactic longitude and 10$h^{-1}$ Mpc in distance. For galactic longitudes ranging from 330$^\circ$ to 30$^\circ$ (masking the bulge) we now consider bins further bound by the lines \mbox{$|b| = 10^\circ$}. Sampling from the adjacent bins (\mbox{$10^\circ < |b| < 20^\circ$}), we populate the bulge with $N$ galaxies drawn at random from the galaxies in adjacent bins; these galaxies are placed at random latitudes and a normal scatter of 20 \kms~is introduced in the velocity to prevent artifacts in the group properties arising due to galaxies at identical redshifts. $N$ is calculated by drawing a random normal deviate from a distribution with a mean equal to the number of galaxies in the two adjacent bins (above and below the plane), then subtracting the number of galaxies already present within the bin. For other galactic longitudes, the latitudes \mbox{$5^\circ < |b| < 15^\circ$} are used to populate the bins with \mbox{$|b| < 5^\circ$}.\footnote{In this case, we set the mean of the normal distribution from which $N$ is drawn to half the number of galaxies in the adjacent bins.}
The catalog, before and after population of the galactic plane, is shown in Figure \ref{Fig:PopulatePlane}. The population of the plane generated an additional 2076 galaxies.
%\placefigone
%\clearpage
%\clearpage
\subsection{Luminosity function} \label{SubSection:LumFunc}
The K-band luminosity function utilized in algorithm is parameterized in terms of a function of the form \citet*{Schechter:1976},
\begin{eqnarray*}
 \Phi(M) =~& 0.4 \ln{(10)} \Phi^\star 10^{0.4(\alpha+1)(M^\star-M)} \\ 
   & \times \exp{[-10^{0.4(M^\star-M)}]}
\end{eqnarray*}
We use the values reported in \citet{Huchra:2005},
\begin{eqnarray} \label{Eq:SchechterFit}
\alpha & = & -1.02 \nonumber \\
M^\star & = & -24.2 \\
\Phi^\star & = & 1.08 \times 10^{-2}~h^3~ \mathrm{Mpc}^{-3} \nonumber
\end{eqnarray}
which have been computed using the galaxies in the 2MRS catalog with galactic latitudes, $|b| > 10^\circ$.

\subsection{Completeness in redshift-space} \label{SubSection:VelocityLimit}
\begin{figure}
\begin{center}
\includegraphics[width=\columnwidth]{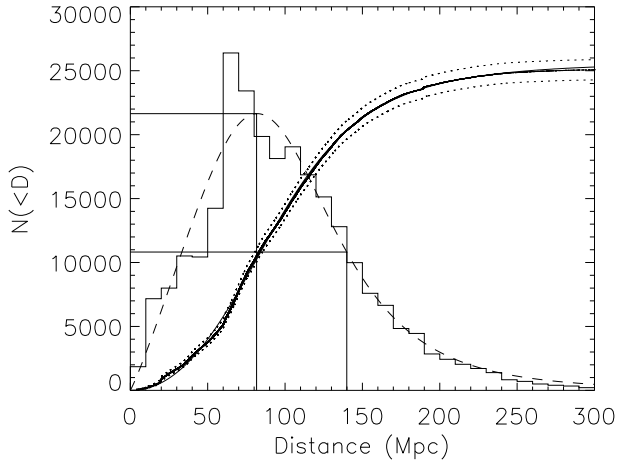}
\end{center}
\caption{\label{Fig:VelLimit}Selection function of the 2MRS survey. The data points (shown as small dots) represent the cumulative number of galaxies as a function of (estimated) distance. The data are fit with a curve of the form of equation (\ref{Eq:VelLimitFit}) using a least-squares fit (solid line). The dotted lines show the 5$\sigma$-errors from Poisson statistics. The data departs from the fit at the both smallest and largest distances shown on the plot. The dashed line shows the derivative of this curve (the selection function) in arbitrary units. The histogram contains the binned data shown in the same arbitrary units as the selection function for purposes of comparison only. The maximum and half-maximum values of the selection function are indicated.
}
\end{figure}

Due to the nature of flux limited surveys, the number density of galaxies observed at sufficiently high redshifts will tend toward zero. At these highest redshifts, the linking lengths used in the algorithm become so large that the majority of identified groups will likely be spurious. For the purposes of building a flow-field model, the groups at the highest redshifts will have the smallest affect on local dynamics, thus we choose to limit the group catalog to a redshift inside which the catalog is reasonably complete.
Figure \ref{Fig:VelLimit} shows the cumulative number of galaxies, $N(<D)$, as a function of (estimated) distance, $D$. The points have been fitted with a curve of the form
\begin{equation} \label{Eq:VelLimitFit}
N(<D) = N_0 \left( \frac{\beta D}{[(\beta D)^b + S^b]^{1/b} } \right)^a
\end{equation}
%
%\placefigtwo
%\clearpage
%\clearpage
%
where the best fit parameters are $N_0 = 2.57 \times 10^4$, $a = 2.10$, $S = 106$ Mpc, $\beta = 0.881$, $b = 3.94$.
The derivative of equation (\ref{Eq:VelLimitFit}) represents the selection function of the survey, $N(D)$, where the number of galaxies observed with estimated distances between $D$ and $D+{\ud}D$ is given by $N(D){\ud}D$. We choose to cut the group catalog at the distance where the selection function falls to half its maximum, \mbox{$D_\mathrm{cut}$ = 140 Mpc}, illustrated in Figure \ref{Fig:VelLimit}. The entire 2MRS $K < 11.25$ dataset plus the galaxies generated in the population of the galactic plane will be used to create the group catalog, but the catalog will then be truncated, excluding groups with mean estimated distances greater than $D_\mathrm{cut}$.

\section{Parameter choices} \label{Section:ApplyAlgorithm}

In this section we justify the choice of linking parameters used in the construction of our group catalog. Any group catalog produced from the remaining data sample will contain minimal biasing effects at the highest and lowest redshifts as well as minimal edge effects across the galactic plane. There remains, however, a choice of the parameters that specify the minimum density contrast of detected groups. These parameters, $D_0$ and $V_0$ from equations (\ref{Eq:DScaling}) and (\ref{Eq:VScaling}) above, must be chosen in a somewhat arbitrary fashion. 

There is no perfect choice of these parameters that will allow us to identify only groups which are gravitationally bound. In any choice we make, some bound systems may be divided and unbound galaxies will be present in some groups. 
For very large values of both $D_0$ and $V_0$, the algorithm will associate all of the galaxies into a single group. Conversely, if we choose sufficiently small $D_0$ or $V_0$, the algorithm will divide substructures within real clusters into multiple systems \citep*{Ramella:1997}, eventually separating each galaxy into its own group. 
It is clear, therefore, that a suitable parameter choice will lie between these extreme cases. 
The method of choosing the specific values of the parameters must still remain arbitrary; in order to be able to infer properties of the Universe (e.g. the matter density parameter) from the catalog it is unwise to calibrate the algorithm using simulations based on a set of defined initial assumptions as this would bias our results towards recovering these initial 
values.\footnote{See Crook et al. \Change{2007a}, \textit{in preparation}, for further investigation.}
It is obvious that there will be a choice of parameters that lie between these extreme values which maximizes the \textit{number} of groups produced. It is therefore reasonable to use a method of maximization to determine the choice of linking parameters, with no alternative method available that does not have similar shortcomings.

\begin{figure*}
\begin{center}
\includegraphics[width=\textwidth]{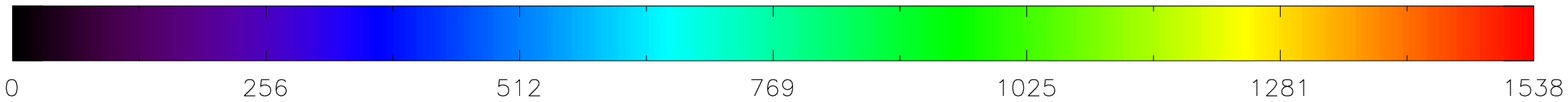} \\
\vspace{5mm}
\includegraphics[width=0.45\textwidth]{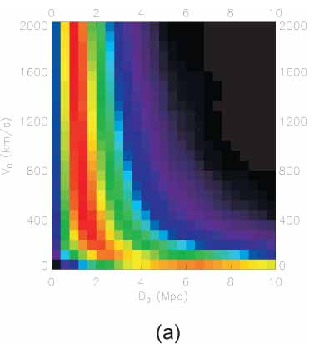}
\hspace{5mm}
\includegraphics[width=0.45\textwidth]{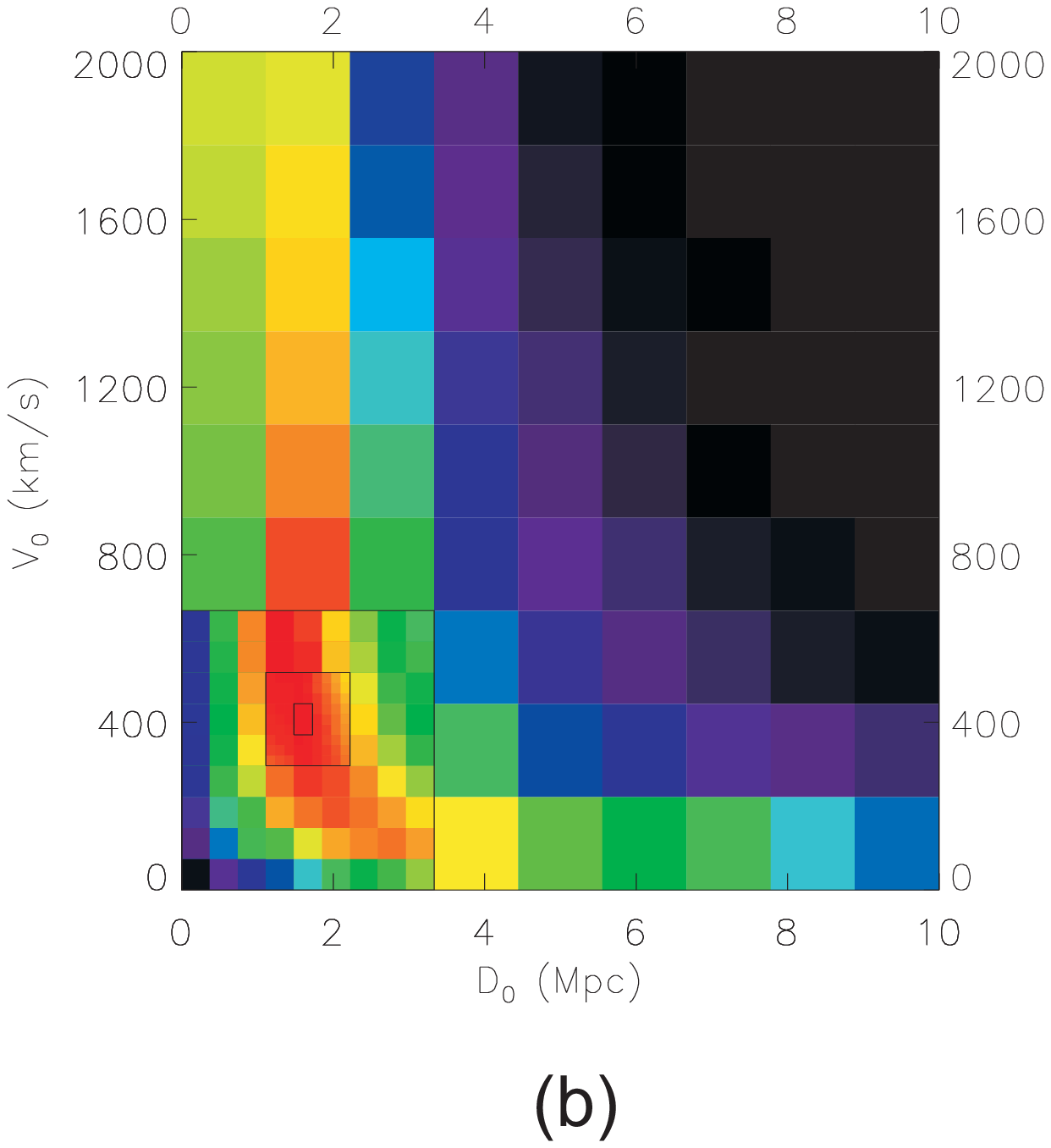}
\end{center}
\caption{\label{Fig:VaryParam}The number of groups of 3 or more galaxies obtained as a function of the parameters $D_0$ and $V_0$. In (a), the clustering algorithm has been executed for each pair of parameters on a \mbox{25$\times$25} grid.
(b) contains a graphical representation of the execution of the maximization routine discussed in the \S\ref{Section:ApplyAlgorithm} of the text.
}
\end{figure*}

At this point, we must consider the size of the group we choose to maximize. We choose to ignore binaries in our definition of groups as previous work has shown such systems identified using percolation algorithms to be unbound in the majority of cases \citep[e.g][]{Diaferio:1999}.
We consider the parameters obtained when maximizing the number of groups of $G$ or more members for $3 \le G \le 20$ as described below.
We choose to set $V_F = 1000$~\kms~following HG82.
Figure \ref{Fig:VaryParam} shows the number of groups containing 3 or more members in $D_0$--$V_0$ space. In Figure \ref{Fig:VaryParam}(a) we explore the parameter space on the intervals \mbox{$D_0$ = [0, 10] Mpc}, \mbox{$V_0$ = [0, 2000] \kms}. We then attempt to maximize the number of groups obtained by the following method: we divide the region spanning 0 $\rightarrow$ 10 Mpc in $D_0$ and 0 $\rightarrow$ 2000 \kms~in $V_0$ into a 9$\times$9 grid and search for the combination of parameters that produces the largest number of groups. We then change the range of the $D_0$ and $V_0$ parameters spanned to coincide with a 3$\times$3 grid (as far as possible) centered on the values of $D_0$ and $V_0$ that produced the largest number of groups. We divide this region into a 9$\times$9 grid and iteratively repeat the procedure until the desired accuracy of the parameters is reached. This procedure is illustrated in Figure \ref{Fig:VaryParam}(b).
%
%\placefigthree
%\clearpage
%\clearpage
%
We repeat this maximization procedure for all values of $G$ between 3 and 20. The obtained value of $D_0$ rises gently with $G$, while the scatter in obtained values of $V_0$ increases rapidly with $G$.
As observed in Figure \ref{Fig:VaryParam}(a), the number of groups depends highly on $D_0$, but is comparatively insensitive to $V_0$. Since the velocity dispersion of a cluster is not expected to depend on the size of the cluster (see the discussion in \S\ref{Section:GroupAlgorithm} above) it is not sensible to consider large values of $V_0$ (i.e. $> 1000~\kms$) as this will introduce members that are not gravitationally bound and thus propagate errors into the mass estimates. The fraction of interlopers present in a group will also increase with both $D_0$ and $V_0$, thus the most sensible choice of parameters to minimize interlopers and reduce the scatter in $V_0$ corresponds to $G=3$.

The maximum number of groups of 3 or more members (\NumGroupsMax) is obtained for the values ($D_0$, $V_0$) = (\DmaxVal~$\pm$ \DmaxErr~Mpc, \VmaxVal~$\pm$ \VmaxErr~\kms), corresponding to the density contrast \mbox{$\delta \rho / \rho$ = \DensityContrastMax}.

In an analysis of the northern CfA redshift survey\footnote{This is a subset of the extended CfA redshift survey \citep{deLapparent:1991,Geller:1989,Huchra:1990,Huchra:1995}.} (hereafter CfAN), \citet{Ramella:1997} show that the group properties are statistically stable for values of density contrasts \mbox{$\delta \rho / \rho \ge 80$}, where they scale $V_L$ in a similar manner to $D_L$ and choose \mbox{$V_0=350$ \kms}. \citet{Diaferio:1999} apply a similar choice of parameters to mock CfA surveys based on $N$-body simulations and conclude that 80\% of groups with 4 or more members are true virialized systems, whereas 40\% of triplets are not, confirming the hypothesis of \citet*{Ramella:1989}. As Figure \ref{Fig:VaryParam}(a) shows minimal variation in the number of groups produced with changing $V_0$ compared to changing $D_0$, the findings of \citet{Ramella:1997} are applicable to this study. We will proceed to analyze the groups produced at both the values of $D_0$ and $V_0$ that maximimize the number of groups of 3 or more members, as well as the values suggested by \citet{Ramella:1997} (i.e. $D_0$ = \Dstd, which corresponds to \mbox{$\delta \rho / \rho = 80$},\footnote{The smallest allowed density contrast is chosen to minimize the probability of splitting the richest systems.} and $V_0$ = \Vstd).

\section{Groups} \label{Section:Groups}

We present the results of applying the group-finding algorithm (\S\ref{Section:GroupAlgorithm}) to the 2MRS catalog subset (\S\ref{Section:DataModifications}) using both pairs of parameters discussed in \S\ref{Section:ApplyAlgorithm} above. The group catalogs are presented in Tables \ref{Tab:GroupCatalogMax}--\ref{Tab:GalaxiesInGroupsStd} in the appendix (see the electronic edition for the complete catalogs). We provide an overview of the catalogs in \S\ref{SubSection:CatalogOverview} below, then discuss the identified groups and contrast the two catalogs in \S\ref{SubSection:Overlap}. We present the properties of the obtained groups (\S\ref{SubSection:GroupProperties}) and discuss the reliability of the clustering algorithm (\S\ref{SubSection:Reliability}).

\subsection{Overview} \label{SubSection:CatalogOverview}
The catalog produced using the parameters \ParamMax~is presented in Table \ref{Tab:GroupCatalogMax}. These parameters produced the maximum number of groups of 3 or more galaxies, and correspond to a density contrast \mbox{$\delta \rho / \rho$ = \DensityContrastMax}; this catalog will hereafter be referred to as the low-density-contrast (LDC) catalog. The catalog produced using the parameters \ParamStd~is presented in Table \ref{Tab:GroupCatalogStd}. These parameters correspond to the density contrast \mbox{$\delta \rho / \rho$ = 80}; this catalog will hereafter be referred to as the high-density-contrast (HDC) catalog.

\begin{figure*}
\begin{center}
\includegraphics[width=\textwidth]{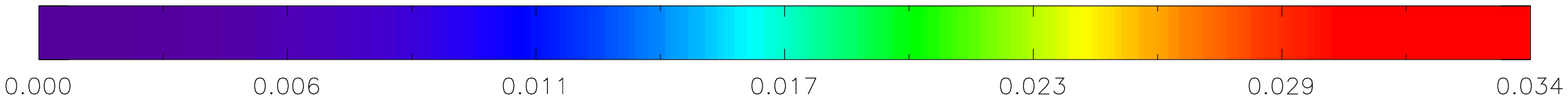} \\
\vspace{5mm}
\includegraphics[width=0.9\textwidth]{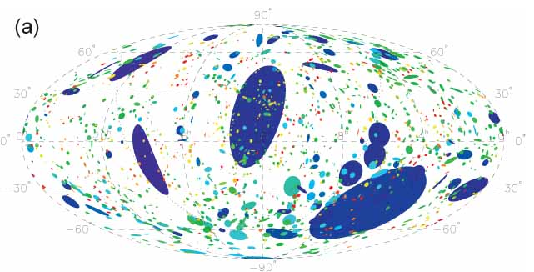} \\
\vspace{5mm}
\includegraphics[width=0.9\textwidth]{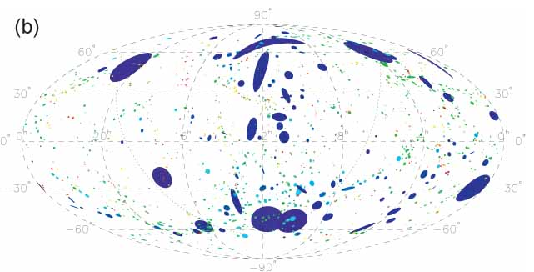}
\end{center}
\caption{\label{Fig:Clusters}Groups (scaled by angular size) identified by clustering algorithm in 2MRS catalog. 
Figures (a) and (b) are shown using a Mollweide projection in equatorial coordinates, centered at 12$^h$00$^m$. Figures (c) and (d) (available only in the electronic edition) are shown using a Mollweide projection in galactic coordinates.
Each group is plotted as an ellipse with the major axis, position angle and axis-ratio of the group. The ellipses are transformed from the $x$-$y$ coordinate system discussed in \S\ref{SubSection:AxisRatio} to the appropriate map projection. The color of the ellipse represents the group's mean redshift. The galactic plane is shown by the dotted line in Figures (a) and (b).
Figures (a) and (c) show the groups in the LDC catalog (\DCMax); the LSC has been clearly identified as the central structure. Figures (b) and (d) show the groups in the HDC catalog (\DCStd); we observe the effect of increasing the minimum density contrast required to identify groups: the LSC has been broken into several constituents.
The distortions created by the map projection are maximized near the poles, and enhanced for larger structures. The seemingly strange shape of the LSC in Figure (c) is the result of mapping an ellipse in the $x$-$y$ coordinate system onto this projection, and is only partly representative of the true shape of the structure.}
\end{figure*}

Figure \ref{Fig:Clusters} shows the positions and sizes of all groups in the two catalogs. Figures \ref{Fig:Clusters}(a) and \ref{Fig:Clusters}(b) show the groups in equatorial coordinates; Figures \ref{Fig:Clusters}(c) and \ref{Fig:Clusters}(d) show the groups in galactic coordinates (available in the electronic edition only).
%
%\placefigfour
%\clearpage
%\thispagestyle{empty}
%\setlength{\voffset}{-25mm}
%\clearpage
%\setlength{\voffset}{0mm}
\begin{figure*}
\begin{center}
\includegraphics[width=\textwidth]{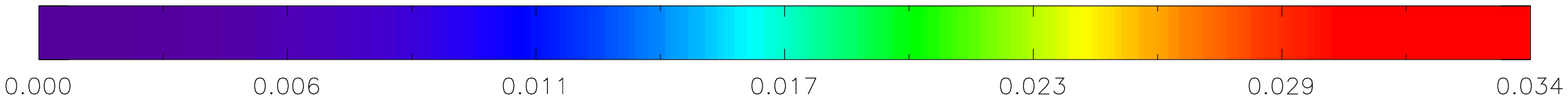} \\
\vspace{5mm}
\includegraphics[width=0.9\textwidth]{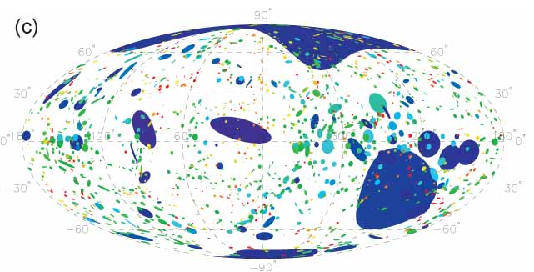} \\
\vspace{5mm}
\includegraphics[width=0.9\textwidth]{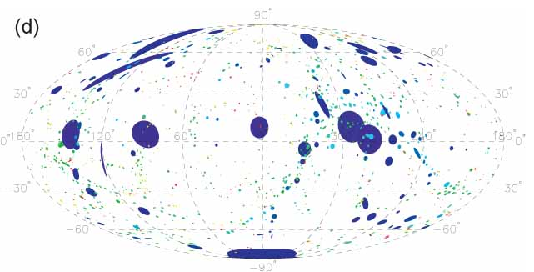} \\
\end{center}
Fig. \ref{Fig:Clusters}.--- \textit{continued...}
\end{figure*}
%\clearpage
%
The maps are shown in Mollweide projection, which preserves the area of the structures on the surface of a sphere but distorts their shape, especially close to the poles. The plots show ellipses that have the properties of the groups discussed in \S\ref{SubSection:AxisRatio} below, and are only representative of the shape and size of the group. The Local SuperCluster (LSC) has been clearly identified in Figure \ref{Fig:Clusters}(a) as the large structure in the center of the figure. When applying the algorithm with the higher minimum density contrast, this structure is split into several constituents as shown in Figure \ref{Fig:Clusters}(b).
(The same result is evident in Figures \ref{Fig:Clusters}(c) and \ref{Fig:Clusters}(d) in the electronic edition, however the LSC encompasses the pole of the coordinate system in this case. The distortions due to the map projection are therefore enhanced in these plots, and the shape is less representative of the true shape of the LSC. The area occupied by the LSC in Figure \ref{Fig:Clusters}(c) is the same as that in Figure \ref{Fig:Clusters}(d). The constituents that have been merged to form the LSC are clearly visible in Figure \ref{Fig:Clusters}(d).)

There is an apparent enhancement in the number of groups with large angular sizes near the galactic plane (see Figures \ref{Fig:Clusters}(b) and \ref{Fig:Clusters}(d), the latter is available in the electronic edition only). There are 5 groups shown as ellipses with major axes greater than 5$^\circ$ and with centers inside $|b| < 10^\circ$. Of these, 2 have only 3 members (of which 2 are genuine galaxies from 2MASS XSC) and only 1 out of the remaining 3 groups has more than 10\% of its members randomly generated in the population of the plane; the remaining 2 are the only groups with more than 5 members, thus the large apparent sizes of all 5 groups in the figure is due to their proximity. We conclude that the observed enhancement is therefore not an artifact of the population of the plane.

\begin{figure*}
\begin{center}
\includegraphics[width=\textwidth]{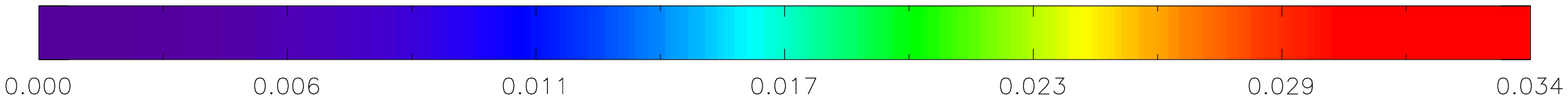} \\
\vspace{5mm}
\includegraphics[width=0.9\textwidth]{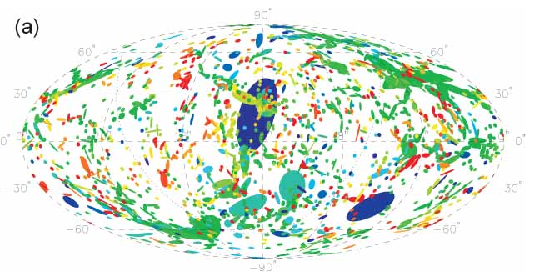} \\
\vspace{5mm}
\includegraphics[width=0.9\textwidth]{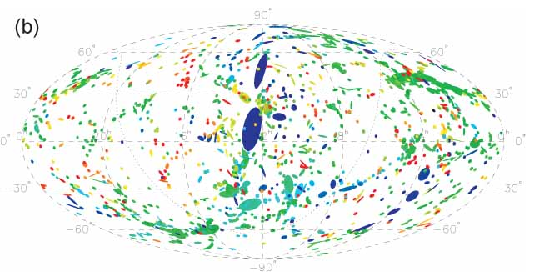}
\end{center}
%\vspace*{-5mm}
\caption{\label{Fig:ClustersN}Groups (scaled by number of members) identified by clustering algorithm in 2MRS catalog shown in equatorial coordinates.
{
Figures (a) and (b) are shown using a Mollweide projection in equatorial coordinates, centered at 12$^h$00$^m$. Figures (c) and (d) (available only in the electronic edition) are shown using a Mollweide projection in galactic coordinates.
Figures (a) and (c) show the groups in the LDC catalog (\DCMax). Figures (b) and (d) show the groups in the HDC catalog (\DCStd).
Each group is plotted as an ellipse with the position angle and axis-ratio of the group. The areas of the ellipses are proportional to the number of members in the group, scaled such that the major-axis of the largest group is 75\% of its true size. The ellipses are transformed from the $x$-$y$ coordinate system discussed in \S\ref{SubSection:AxisRatio} to the appropriate map projection.
The color of the ellipse represents the group's mean redshift.
The galactic plane is shown by the dotted line in Figures (a) and (b).
}
}
\end{figure*}

Figure \ref{Fig:ClustersN} shows the same groups as Figure \ref{Fig:Clusters}, however the area of each ellipse is proportional to the number of members in the group, rather than the square of the group's angular size.
%
%\placefigfive
%\clearpage
\begin{figure*}
\begin{center}
\includegraphics[width=\textwidth]{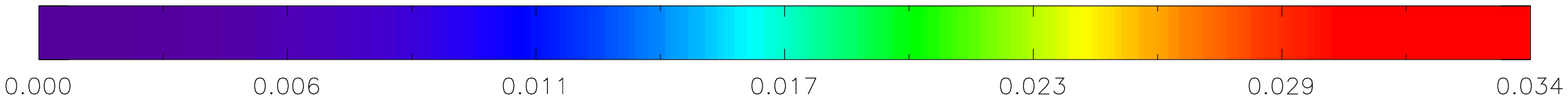} \\
\vspace{5mm}
\includegraphics[width=0.9\textwidth]{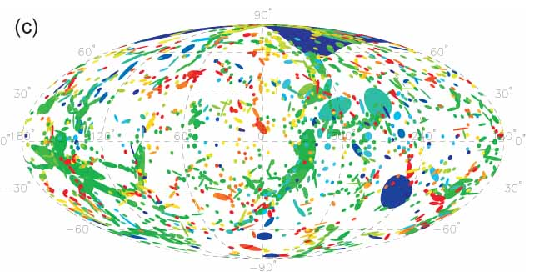} \\
\vspace{5mm}
\includegraphics[width=0.9\textwidth]{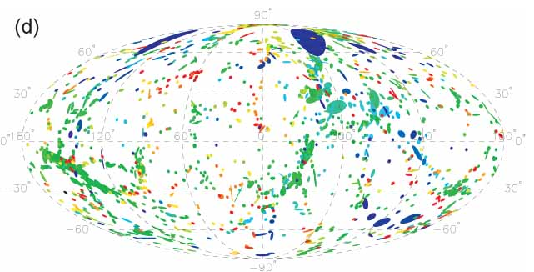} \\
\end{center}
Fig. \ref{Fig:ClustersN}.--- \textit{continued...}
\end{figure*}
%\clearpage
%
The areas are normalized such that the major-axis (see \S\ref{SubSection:AxisRatio}) of the largest group appears as 75\% of its true angular size. These figures are no longer dominated by the foreground groups that have the largest angular sizes, but show how the groups are distributed across the sky out to the redshift limit of the sample.

\subsection{Identification and overlap} \label{SubSection:Overlap}
\begin{figure*}
\begin{center}
\includegraphics[width=0.85\textwidth]{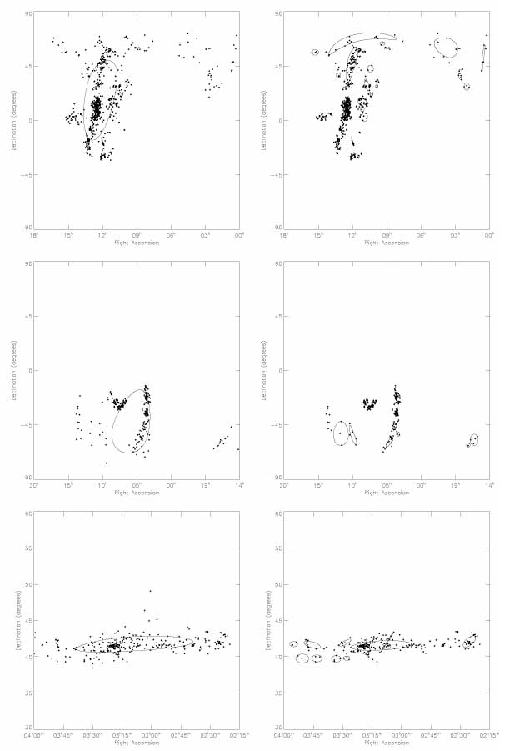}
\end{center}
\caption{\label{Fig:BigGroups}Six largest groups in the LDC catalog (\DCMax). 
The galaxies in these groups are shown on the plots on the left hand side. The groups identified in the HDC catalog (\DCStd) in the same region of the sky are shown on the right hand side. The ellipses shown have been computed in the $x$--$y$ coordinate space discussed in \S\ref{SubSection:AxisRatio} and mapped onto the equatorial coordinates used in the figure. From top to bottom, the identifications of groups with known clusters and superclusters is as follows --- 1st page: Top: Virgo, NGC3607, NGC4105, IC764, NGC5746, NGC3190, NGC5846 and NGC4038 Clusters, M81 group (LDC Group {\#}852). Note that the ellipse on the right-hand figure at (13h, +10$^\circ$) is partially masked by the high density of points. Middle: Eridanus, Fornax I, Dorado, NGC2280, NGC1433, NGC2559 Clusters (LDC Group {\#}391). Bottom: Perseus-Pisces (A426, A347) (LDC Group {\#}229). 2nd page: Top: Hydra (A1060) (LDC Group {\#}712). Middle: Norma (A3627, the GA) (LDC Group {\#}1117). Bottom: Centaurus (A3526) (LDC Group {\#}881).}
\end{figure*}

Due to the the nature of the group finding algorithm, we expect to find that all of the galaxies assigned to groups in the HDC catalog will also be assigned to groups in the LDC catalog, however the converse is not necessarily true. We consider all the galaxies assigned to groups in the HDC catalog and determine the correspondence between groups in the two catalogs. The six largest groups in the LDC catalog are plotted in Figure \ref{Fig:BigGroups}; the corresponding groups in the HDC catalog are also shown. The largest group in the LDC catalog contains 810 members (of which 2 were randomly generated in the population of the galactic plane). This group is the result of merging several smaller groups including Virgo, NGC3607, NGC4105, IC764, NGC5746, NGC3190, NGC5846, and NGC4038 Clusters, as well as the M81 group, and corresponds to the LSC. In the HDC catalog, most of these groups have been identified individually; in fact Virgo has been split into two groups, containing 298 galaxies and 123 galaxies, respectively. 
Eridanus, Fornax I, Dorado and NGC2280, NGC1433, NGC2559 Clusters merge to form the second largest group in the LDC catalog (302 members, including 22 simulated); again these were identified individually in the HDC catalog.
A426 and A347 of the Perseus-Pisces supercluster make up the third largest group in the LDC catalog, containing 301 galaxies, but these were identified as two separate groups in the HDC catalog.
Hydra (A1060) is identified as the fourth largest with 241 members,
and Norma (A3627, the GA) is identified as the fifth with 217 members (42 of which were randomly generated).
Centaurus (A3526) was identified as the sixth largest group (202 members).
This exercise demonstrates that the correct choice of parameters used in the group algorithm is entirely dependent on the size of the structures that are desired. 

The remaining identifications and correspondence between the catalogs are shown in the appendix (Tables \ref{Tab:GroupCatalogBigMax} and \ref{Tab:GroupCatalogBigStd}), where we consider only groups containing 25 or more members in the HDC catalog.
Tables \ref{Tab:GalaxiesInGroupsMax} and \ref{Tab:GalaxiesInGroupsStd} contrast the group assignments of individual galaxies between the two catalogs.
It may be surprising that the parameters chosen to maximize the total number of groups actually merge several of the large groups, hence apparently reducing the total number of groups. Although not obvious, this result is not unexpected because the larger linking length will allow many smaller groups to be identified that do not exist in the HDC catalog. The latter association is generating more groups than are removed by the merging of largest groups.
It is likely that the groups in the LDC catalog contain a higher fraction of interlopers than the groups in the HDC catalog, however it is evident that the LDC catalog identifies the largest structures on the sky. This suggests that the LDC catalog will be the better candidate for the basis of a flow-field model, as some of the largest structures are fragmented in the HDC catalog. This is explored further in 
% In prep reference
follow-up work.\footnote{See Crook et al. \Change{(2007b)}, \textit{in preparation}.}
%
%\placefigsix
%\clearpage
%\thispagestyle{empty}
%\setlength{\voffset}{-25mm}
%\clearpage
%\setlength{\voffset}{0mm}
\begin{figure*}
\begin{center}
\includegraphics[width=0.85\textwidth]{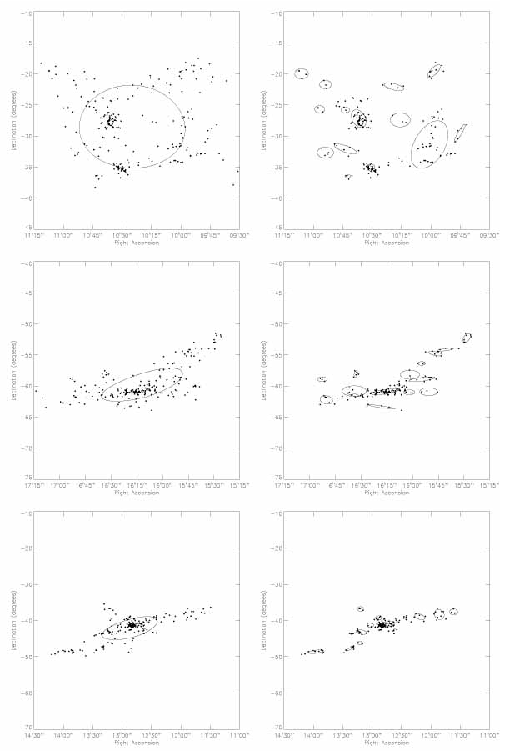} \\
\end{center}
Fig. \ref{Fig:BigGroups}.--- \textit{continued...}
\end{figure*}
%\clearpage
%

\subsection{Group properties} \label{SubSection:GroupProperties}

In this section, we discuss the properties of the LDC and HDC catalogs. Estimates of the velocity dispersion, size, mass and luminosity are discussed in \S\ref{SubSection:BasicGroupProperties}; we obtain estimates of the axis-ratio and position angle of the groups in \S\ref{SubSection:AxisRatio} below.

\subsubsection{Basic properties} \label{SubSection:BasicGroupProperties}

The properties of the LDC and HDC catalogs are summarized in Table \ref{Tab:CatalogProperties}.
In further analysis, we only consider groups with 5 or more members that are also present in the 2MRS catalog (referred to as \textit{genuine} hereafter), as opposed to those generated in the population of the galactic plane. We do this in an attempt to exclude groups with a high fraction of interlopers in our analysis.
%
%\clearpage
% Table 1
\begin{table*} 
\caption{\label{Tab:CatalogProperties}General properties of the group catalogs}
\begin{center}
\begin{tabular}{ccc}
\tableline\tableline
Property                        & LDC Catalog   & HDC Catalog \\
\tableline
$\delta \rho / \rho$            & \DensityContrastMax  & 80 \\
D0 (Mpc)                        & \DmaxVal      & \DstdVal \\
V0 (\kms)                       & \VmaxVal      & \VstdVal \\
No. of Singles                  & 5548 (27.2\%) & 9608 (47.2\%) \\
No. of Binaries                 & 1397 (13.7\%) & 1710 (16.8\%) \\
No. of Groups of 3+             & 1538 (59.1\%) & 1258 (36.1\%) \\
No. of Groups of 10+            & 203 (30.5\%)  & 113 (13.3\%) \\
No. of Groups of 50+            & 17 (10.7\%)   & 8 (4.7\%) \\
Mean no. per group              & 7.84          & 5.84 \\ 
Standard Deviation              & 26.25         & 11.85 \\
Min/Max per group               & 3 / 810       & 3 / 298 \\
\tableline
\end{tabular}
\end{center}
\tablecomments{The values in parentheses represent the percentages of galaxies that fall into this category.}
\end{table*}

%\clearpage
%
We provide two estimates of the mass of the groups. We first compute the virial mass of the group, $M_V$,
\begin{equation}
M_\mathrm{V} = \frac{3\pi}{2} \frac{\sum_\mathrm{P}^2 R_\mathrm{P}}{G}
\end{equation}
where $G$ is Newton's constant, $\sigma_\mathrm{P}$ is the projected velocity dispersion,
\begin{equation}
\sigma_\mathrm{P}^2=\frac{\sum_i (V_i - V_G)^2}{N-1}
\end{equation}
and $R_\mathrm{P}$ is the projected virial radius,
\begin{equation}
R_\mathrm{P}=\frac{N(N-1)}{\sum_{i>j} R_{ij}^{-1}}
\end{equation}
$V_G$ is the mean group velocity, and $V_i$ is the line-of-sight velocity of the $i^\mathrm{th}$ member. $N$ is the total number of galaxies in the group and $R_{ij}$ is the projected separation between two galaxies, defined in terms of their angular separation, $\theta_{ij}$, through
\begin{displaymath}
R_{ij}=\frac{2V_G}{H_0} \tan \left( \frac{\theta_{ij}}{2} \right) 
\end{displaymath}
Due to the biases in the virial mass estimator \citep[e.g.][]{Bahcall:1981}, we also calculate the projected mass estimator, $M_P$ \citep{Bahcall:1981,Heisler:1985},
\begin{equation}
M_\mathrm{P} = \frac{f_\mathrm{PM}}{\pi G (N-\gamma)} \sum_i s_i (V_i - V_G)^2 
\end{equation}
where $s_i$ is the offset of the $i^\mathrm{th}$ member (in physical units) from the center of the group. Following \citet{Heisler:1985}, we set $\gamma=1.5$ and $f_\mathrm{PM}=10.2$.

We estimate the total \Change{isophotal} K-band luminosity, $L_K$ of each cluster from the observed luminosity using equation (\ref{Eq:TotalL}) below.
\begin{equation} \label{Eq:TotalL}
L_K = \left[ 1-\frac{\gamma(\alpha+2,L_\mathrm{lim} / L^\star)}{\Gamma(\alpha+2)} \right]^{-1} L_\mathrm{obs} 
\end{equation}
where $L_\mathrm{obs}$ is the total \Change{isophotal} observed K-band luminosity and $L_\mathrm{lim}$ is the limiting observable luminosity at the distance of the cluster, $d_\mathrm{c}$,
\begin{equation}
L_\mathrm{lim}=10^{0.4(M_{\odot,\mathrm{K}}-m_\mathrm{lim}+25+5\log(d_\mathrm{c} / \mathrm{Mpc}))} \Lsun
\end{equation}
$\alpha$ and $L^\star$ take the values used quoted in equation (\ref{Eq:SchechterFit}) above, given
\begin{displaymath}
L^\star=10^{0.4 \log (M_{\odot,\mathrm{K}}-M^\star)} \Lsun
\end{displaymath}
and $\gamma(m, x)$ is the lower incomplete gamma function,
\begin{displaymath}
\gamma(m, x) = \int_0^x t^{m-1} e^{-t} \ud t
\end{displaymath}
and
\begin{displaymath}
\Gamma(m) = \gamma(m, \infty)
\end{displaymath}
\Change{We set the K-band magnitude zero point, $M_{\odot,\mathrm{K}} = 3.39$ \citep{Johnson:1966}. Note that we have not applied an isophotal correction to the luminosities, thus the luminosities presented in this paper are lower than the total K-band luminosities.}

Table \ref{Tab:GroupProperties} contains the median properties of the groups in the catalog that have at least 5 genuine members.
%
%\clearpage
% Table 2
\begin{table*}
\caption{\label{Tab:GroupProperties}Median properties of groups with 5 or more genuine members}
\begin{center}
\begin{tabular}{ccc}
\tableline\tableline
Property                           & LDC Catalog          & HDC Catalog \\
\tableline
$\sigma_\mathrm{P}$ (\kms)         & 197 (183, 206)       & 183 (166, 193)       \\
$R_\mathrm{PV}$ (Mpc)              & 1.71 (1.58, 1.85)    & 0.97 (0.90, 1.04)    \\
$\log[M_V/ \Msun]$                 & 13.79 (13.72, 13.90) & 13.54 (13.46, 13.60) \\
$\log[M_P/ \Msun]$                 & 14.05 (13.98, 14.10) & 13.66 (13.57, 13.75) \\
$\log[(M_V/L_K)/(\Msun / \Lsun)]$  & 1.66 (1.59, 1.72)    & 1.49 (1.41, 1.55)    \\
$\log[(M_P/L_K)/(\Msun / \Lsun)]$  & 1.86 (1.81, 1.93)    & 1.63 (1.56, 1.69)    \\
$\Omega_{M,V}$                     & 0.14 (0.12, 0.16)    & 0.10 (0.08, 0.11)    \\
$\Omega_{M,P}$                     & 0.23 (0.20, 0.27)    & 0.13 (0.11, 0.16)    \\
\tableline
\end{tabular}
\tablecomments{The median values are shown, with 99\% confidence levels in parentheses. We compute the confidence levels by drawing an equally-sized sample from the observed distributions and computing the median values 5,000 times.}
\end{center}
\end{table*}

\begin{figure*}
\begin{center}
\includegraphics[width=0.45\textwidth]{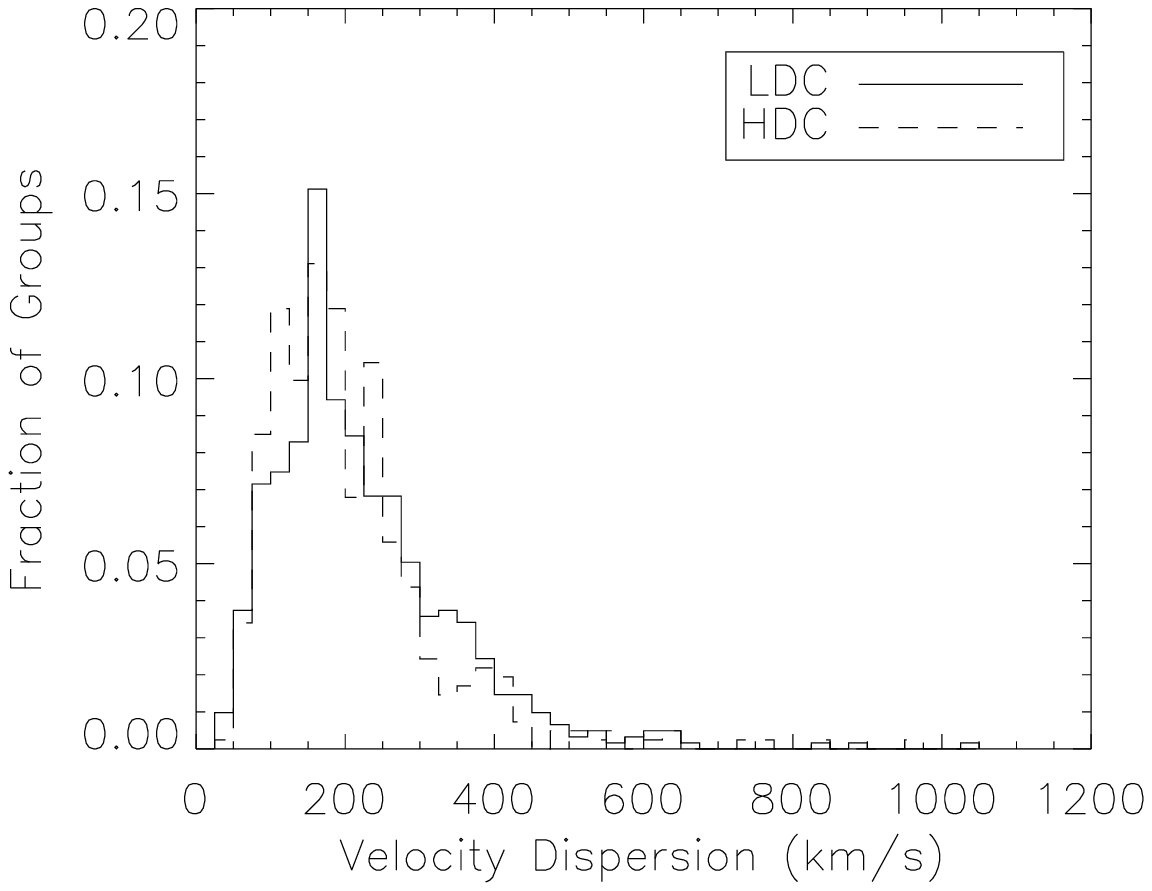}
\hspace{5mm}
\includegraphics[width=0.45\textwidth]{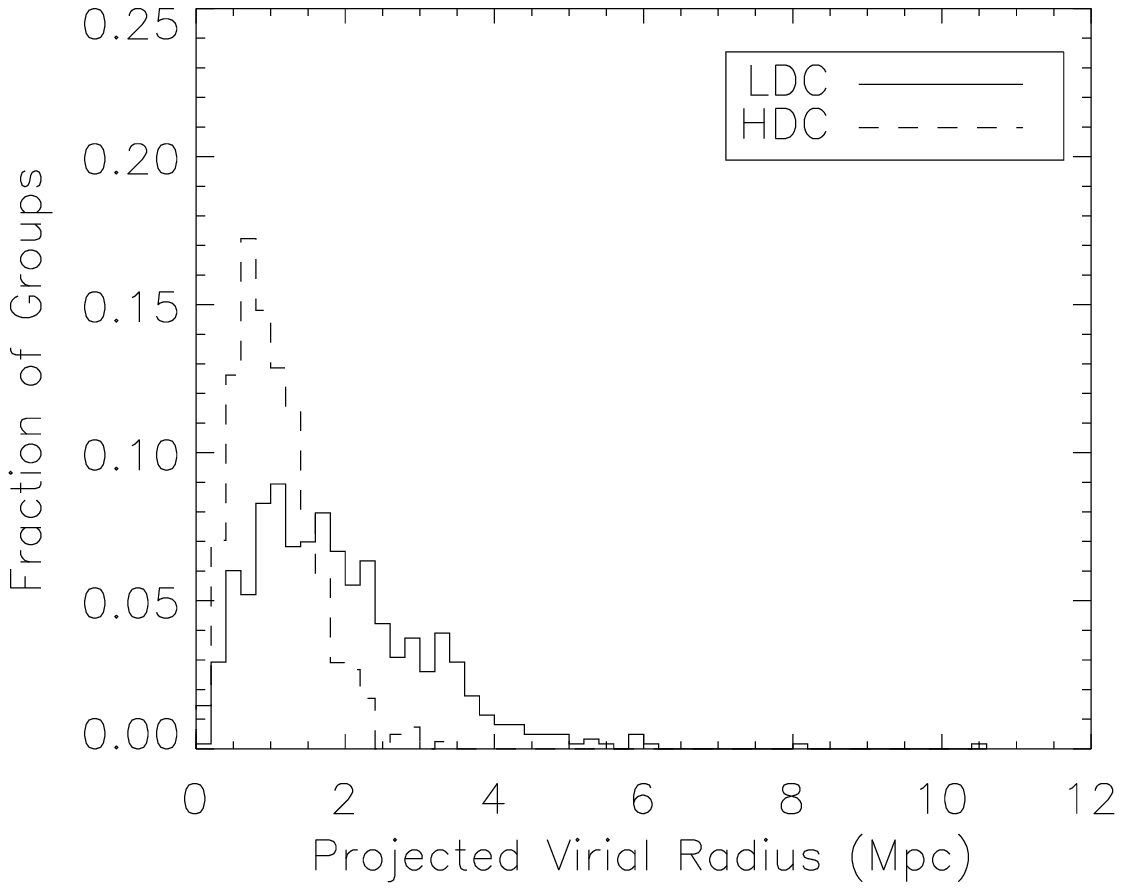}
\vspace{5mm}
\includegraphics[width=0.45\textwidth]{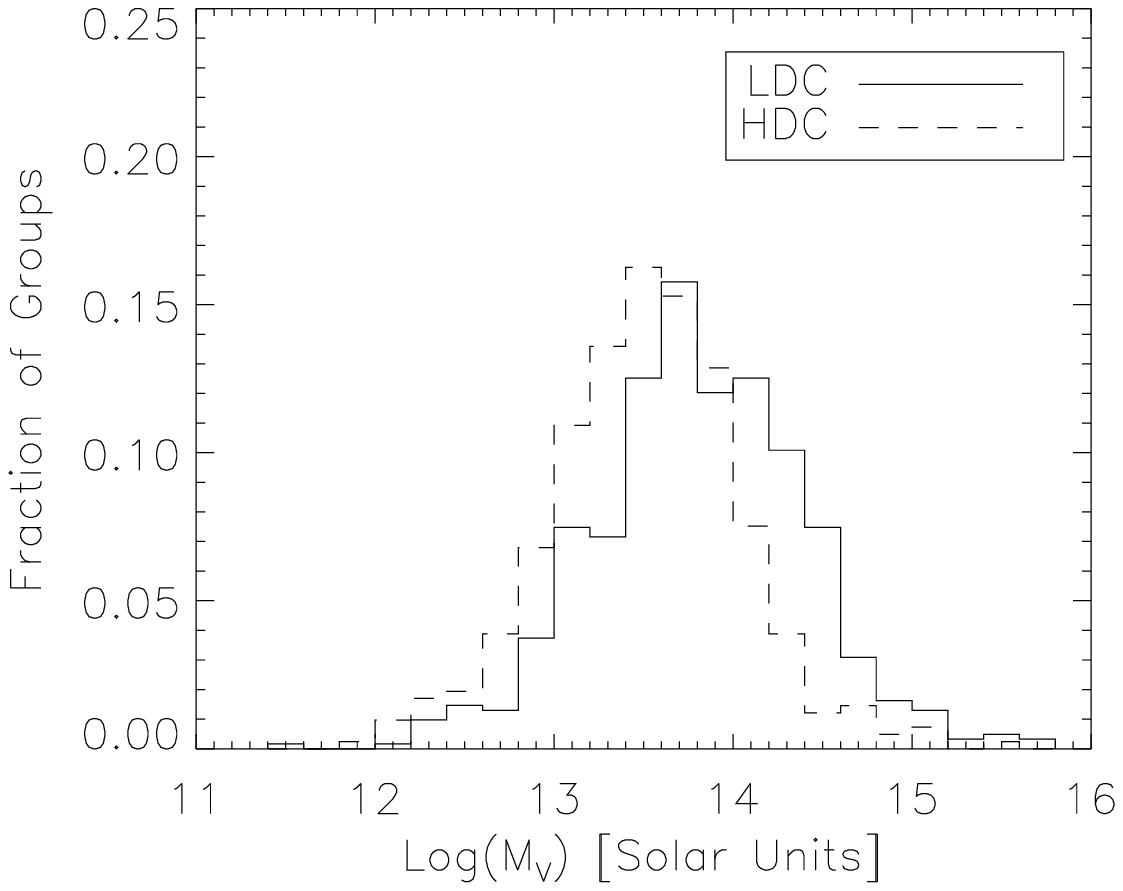}
\hspace{5mm}
\includegraphics[width=0.45\textwidth]{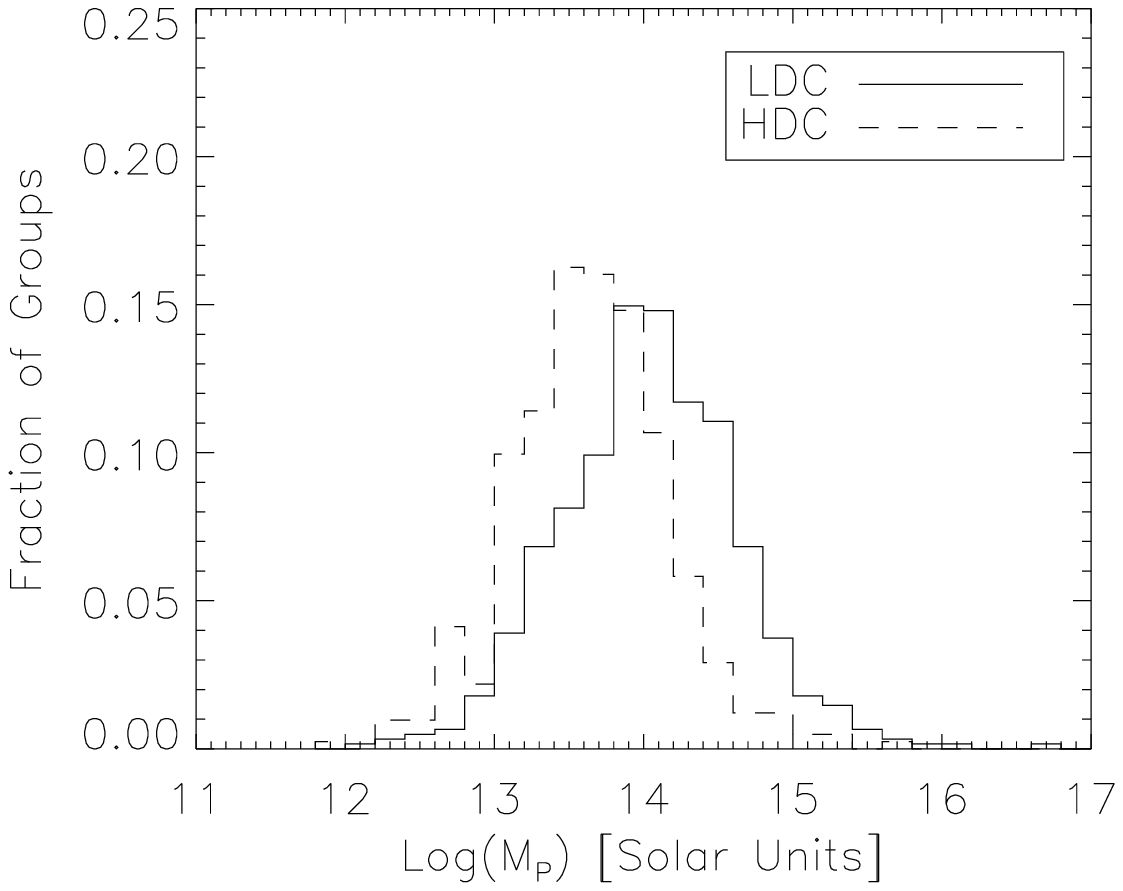}
\vspace{5mm}
\includegraphics[width=0.45\textwidth]{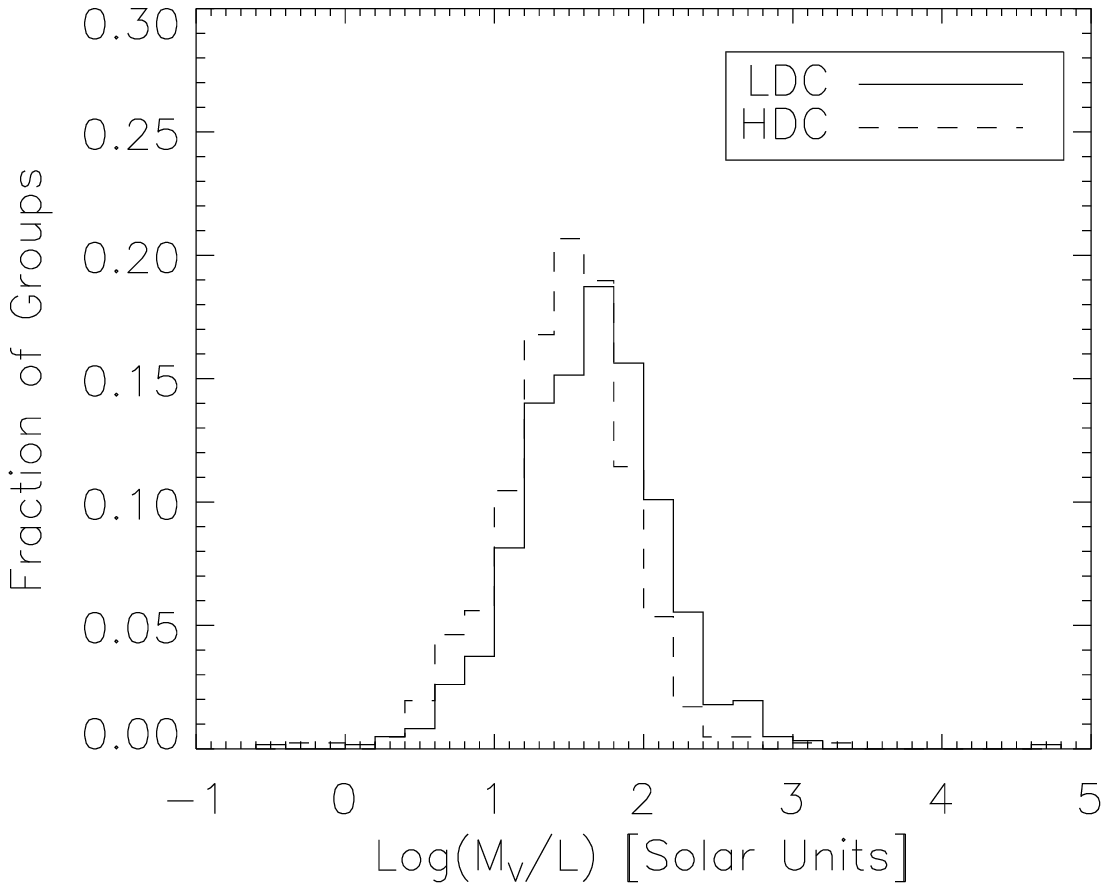}
\hspace{5mm}
\includegraphics[width=0.45\textwidth]{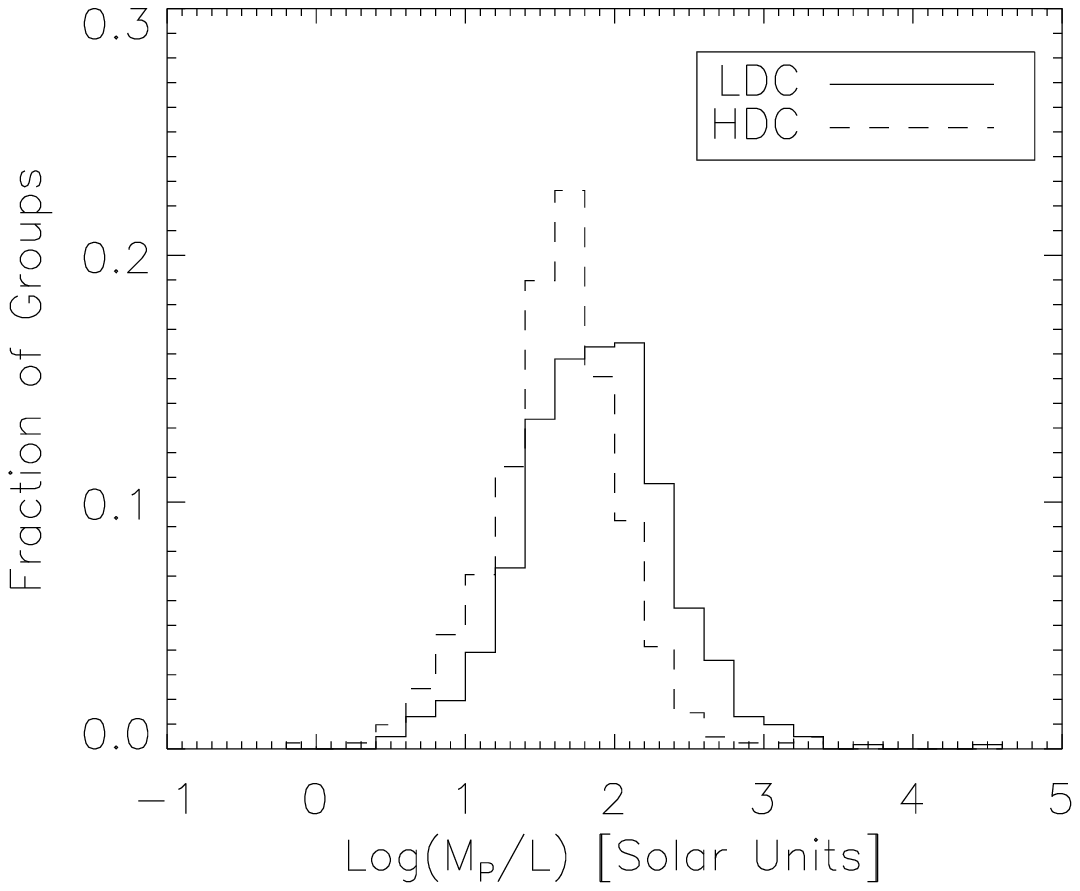}
\end{center}
\caption{\label{Fig:GroupProperties}Properties of groups with at least 5 genuine members. 
{
We contrast the properties of the groups in the LDC catalog (solid line) with those in the HDC catalog (dashed line). Top left: variation in line-of-sight velocity dispersion across the groups. Top right: shows the projected virial radius of each of the groups. Middle: Virial (left) and projected (right) mass estimates of the groups. Bottom: Mass-to-light ratios of the groups using the virial (left) and projected (right) mass estimators. \HAssumption
}
}
\end{figure*}

%\clearpage
%
Figure \ref{Fig:GroupProperties} shows the fraction of groups, containing at least 5 genuine members, as a function of velocity dispersion, projected virial radius, mass and mass-to-light ratio (using both virial and projected mass estimates). 
%
%\placefigseven
%\clearpage
%\clearpage
%
As expected, the distribution of virial radii widens and increases with the larger choice of $D_0$. Similarly, the velocity dispersions increase with the larger choice of $D_0$ and $V_0$ due principally to the change in $V_0$; re-calculating the groups for various values of $D_0$ while keep $V_0$ constant has little impact on the distribution of velocity dispersions. The net effect is expected to increase the estimated virial mass of the groups, as observed.

We use the median mass-to-light ratio to obtain a value of $\Omega_m$, the ratio of the matter-density, $\rho_m$ (assuming all the mass is contained within galaxy clusters) to the critical density, $\rho_c$,
\begin{displaymath}
\rho_c = \frac{3 H_0^2}{8 \pi G}
\end{displaymath}
We use the K-band luminosity density, $\mathcal{L}_K$,
\begin{equation}
\mathcal{L}_K = \Phi^\star L^\star \Gamma(\alpha+2)
\end{equation}
where $\alpha$, $L^\star$ and $\Phi^\star$ take the values given in equation (\ref{Eq:SchechterFit}), in conjunction with the median mass-to-light ratios to estimate $\rho_m$,
\begin{equation}
\rho_{m,E} = \langle \frac{M_E}{L_K} \rangle \cdot \mathcal{L}_K
\end{equation}
where $E$ refers to the method of mass estimation (i.e. \textbf{V}irial or \textbf{P}rojected). \Change{For the luminosity function and magnitude zero point used in this paper, the critical density corresponds to $\langle {M_E}/{L_K} \rangle = 322 \Msun / \Lsun$. Similar results were found by \citet{Kochanek:2001} and \citet{Bell:2003}, adjusting for the authors' isophotal corrections where appropriate.}

The obtained values are given in Table \ref{Tab:GroupProperties}.
\citet{Spergel:2006} obtain $\Omega_m = 0.238^{+0.013}_{-0.024}$ assuming \hvalue; this value only agrees, at the 1$\sigma$-level, with the value we obtained using the projected mass estimator in the LDC catalog ($\Omega_m = 0.229_{-0.012}^{+0.016}$, 1$\sigma$ errors). The virial mass estimates predict $\Omega_m$ to be too small. For this reason, we will use the projected mass estimator as opposed to the virial mass estimator in further analysis. The fact that the HDC catalog predicts a value of $\Omega_m$ significantly smaller than that obtained using WMAP suggests that we are missing a significant fraction of the mass of the cluster in our estimate; selecting groups based on the density contrast $\delta \rho / \rho = 80$ is causing us to underestimate the median mass-to-light ratio of groups. This suggests that the dark matter halos extend beyond the $\delta \rho / \rho = 80$ density-contrast contour that was inferred from luminous matter.

\subsubsection{Orientation and ellipticity} \label{SubSection:AxisRatio}
We include in the catalog a measure of the axis-ratio, position angle and semi-major axis of the groups containing 5 or more genuine members, calculated using the following method: We rotate the coordinates such that center of the group\footnote{We define the center as the mean position of the galaxies in cartesian coordinates, assuming the galaxies lie on the surface of a unit sphere.} lies along the $z$-axis. We measure the angle of each galaxy from the $z$-axis, $\Theta$, as well as its azimuthal angle $\psi$, then define
\begin{eqnarray}
x & = & \Theta \cos(\psi) \\
y & = & \Theta \sin(\psi) 
\end{eqnarray}
such that the positive $y$-axis points north and the positive $x$-axis points east.\footnote{Note that the mean values of $x$ and $y$ are not strictly zero but, in the analyzed data, are sufficiently small that an iterative centering procedure is not required.} This definition is chosen such that the shape of the group is not distorted under the projection onto a plane.

\begin{figure}
\begin{center}
\includegraphics[width=\columnwidth]{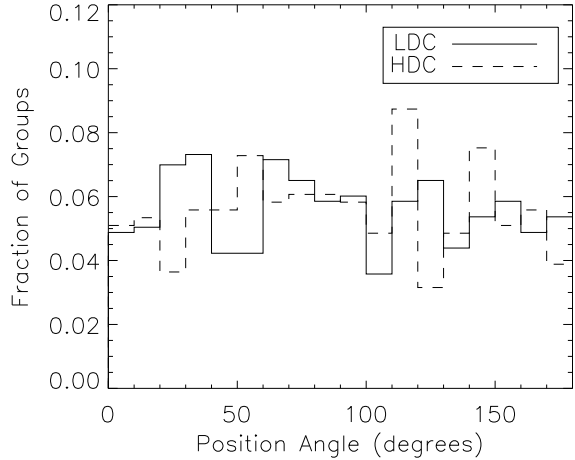}
\end{center}
\caption{\label{Fig:PosAngle}Position angles of groups with 5 or more genuine members. The figure shows the fraction of groups with position angles in the specified intervals (binned by 10$^\circ$). The groups in the LDC catalog are represented by the solid line, and the HDC catalog by the dashed line.}
\end{figure}

We rotate the axes to some angle, $\phi$, and define the coordinates of the galaxies in the rotated frame as ($\tilde{x}_i$, $\tilde{y}_i$). We choose the value of $\phi$ that minimizes $\Sigma_i \tilde{y}_i^2$. We compute the 75$^\mathrm{th}$--percentile values of $|\tilde{x}|$ and $|\tilde{y}|$ ($\tilde{x}_{75}$ and $\tilde{y}_{75}$) and record their ratio, $\eta$, ($0 < \eta < 1$) as a measure of the axis-ratio of the group. We also record the larger of $\tilde{x}_{75}$ and $\tilde{y}_{75}$ as a measure of the semi-major axis of the cluster, $a$, as well as the angle of rotation of the semi-major axis from north toward east (the position angle of the group, $\phi$). We verify that these angles are approximately uniformly distributed by showing the number of galaxies as a function of position angle in Figure \ref{Fig:PosAngle}.
%
%\placefigeight
%\clearpage
%\clearpage
%

The above properties of the groups are illustrated graphically, by using ellipses with the same semi-major axes, axis-ratios and position angles in Figure \ref{Fig:Clusters} and with the same axis-ratios and position angles in Figure \ref{Fig:ClustersN}. For groups with 3--4 members, a circle is drawn with an angular radius equal to the 75$^\mathrm{th}$--percentile mean offset.\footnote{This value is also reported under the column titled $a$ in Tables \ref{Tab:GroupCatalogMax} and \ref{Tab:GroupCatalogStd}.}

In Figure \ref{Fig:BigGroups}, we show the galaxies that are associated by the clustering algorithm to form the six largest groups in the LDC catalog. The corresponding ellipses have been overlayed on this plot.\footnote{Note that the shapes of the ellipses have been distorted due to the choice of coordinate system.} The figure also shows the corresponding groups identified in the HDC catalog. It is evident that the higher density contrast used in the latter choice of parameters splits the large structures identified when choosing a lower-density contrast.

\subsection{Reliability of the algorithm} \label{SubSection:Reliability}

In this section we discuss the verifications performed to ensure that the groups obtained are consistent with both expectation and the literature.
In \ref{SubSection:DistanceDependence} we examine the distance-dependence of the velocity dispersions and mass-to-light ratios of the groups.
We compute the mass functions of the group catalogs and compare them with expectation in \ref{SubSection:MassFunctions}. Finally, we compare the 2MRS group catalogs directly with the UZC-SSRS2 and CfAN group catalogs in \ref{SubSection:OtherCatalogs}.

\subsubsection{Variation with distance} \label{SubSection:DistanceDependence}

\begin{figure*}
\begin{center}
\includegraphics[width=0.45\textwidth]{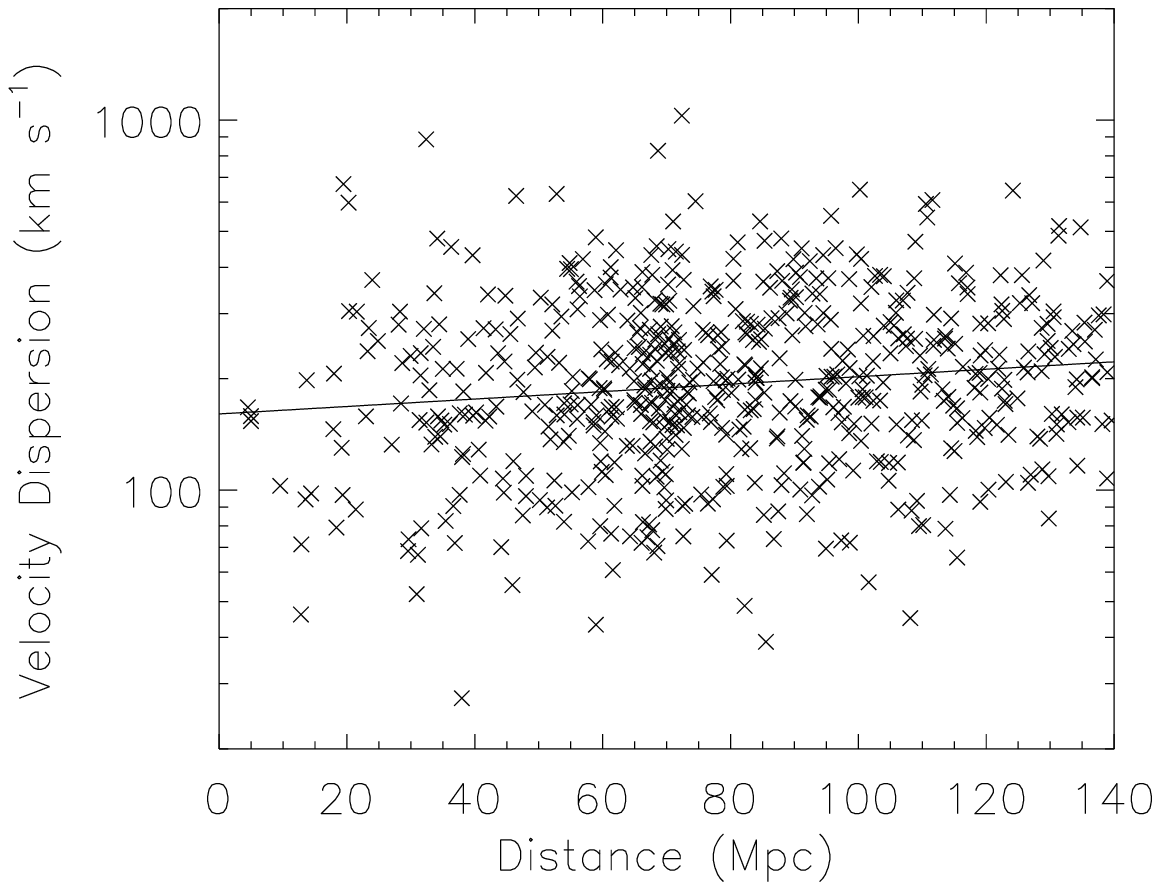}
\hspace{5mm}
\includegraphics[width=0.45\textwidth]{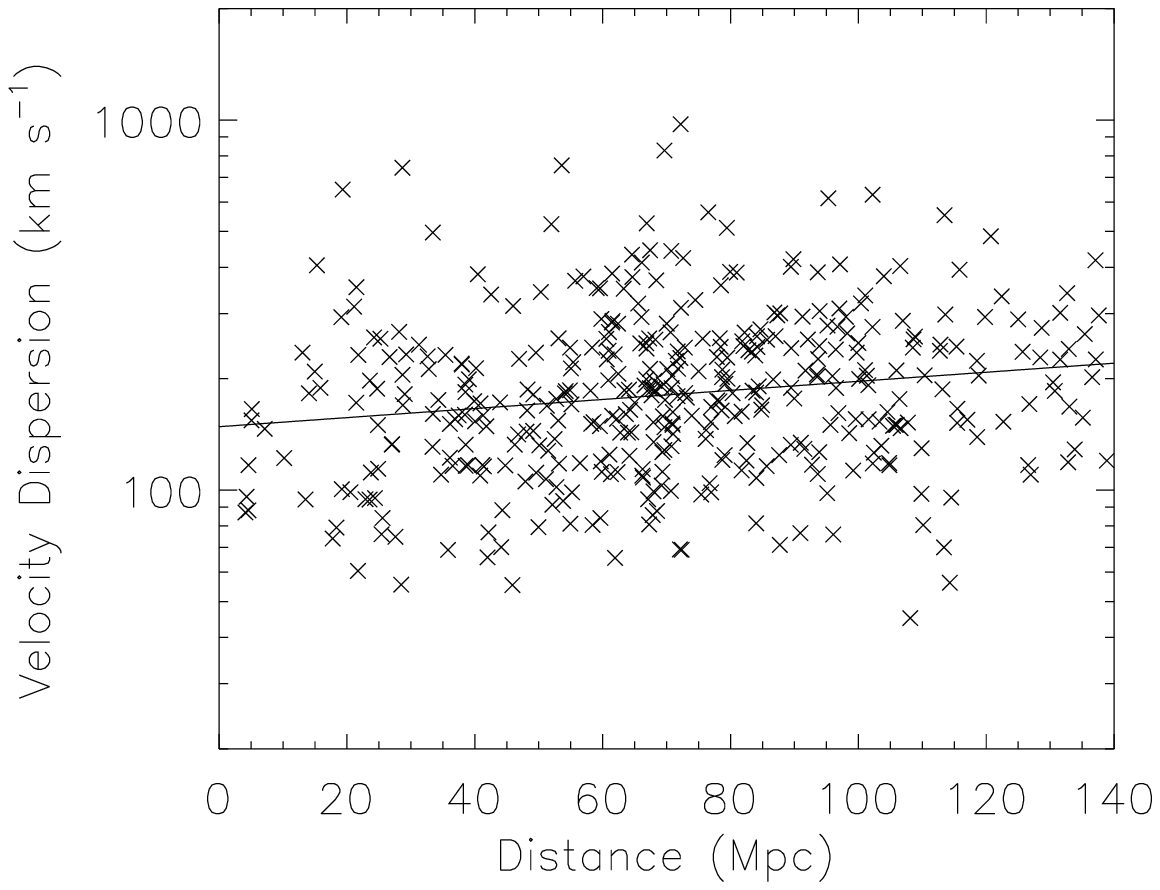}
\end{center}
\caption{\label{Fig:VelocityDispersion}Velocity dispersion of groups as a function of distance. The left panel shows the groups in the LDC catalog (\DCMax). The right panel shows the groups in the HDC catalog (\DCStd). Only groups with 5 or more genuine members are included. The solid line shows a linear fit to the data as described in the text.}
\end{figure*}

Figure \ref{Fig:VelocityDispersion} shows the velocity dispersion, $\sigma_P$, of the groups as a function of distance.
%
%\placefignine
%\clearpage
%\clearpage
%
We fit a curve of the form
\begin{equation} \label{Eq:VDFit}
\sigma_P = 10^{(\alpha D + \beta)}~\mathrm{km~s}^{-1}
\end{equation}
to the data, where the best-fit parameters are given in Table \ref{Tab:BestFitMLVD}.
The large scatter and very small correlation observed in Figure \ref{Fig:VelocityDispersion} (the mean velocity dispersion changes by a factor of 0.63$\bar{\sigma}$ between [20, 40] Mpc and [120, 140] Mpc in the HDC catalog, and 0.35$\bar{\sigma}$ in the LDC 
catalog\footnote{$\bar{\sigma}$ represents the average standard deviation, weighting the standard deviation of the velocity dispersions in each interval equally.})
demonstrates that there is minimal bias introduced in the velocity dispersion of groups with distance; this was desired in the construction of the algorithm (see \S\ref{Section:GroupAlgorithm}). Had we chosen to scale $V_0$ with distance, we would expect the velocity dispersions of the most distant groups to be larger than observed in this figure. Since the correlation is already slightly positive, scaling $V_0$ would have introduced a more significant bias with distance.

\begin{figure*}
\begin{center}
\includegraphics[width=0.45\textwidth]{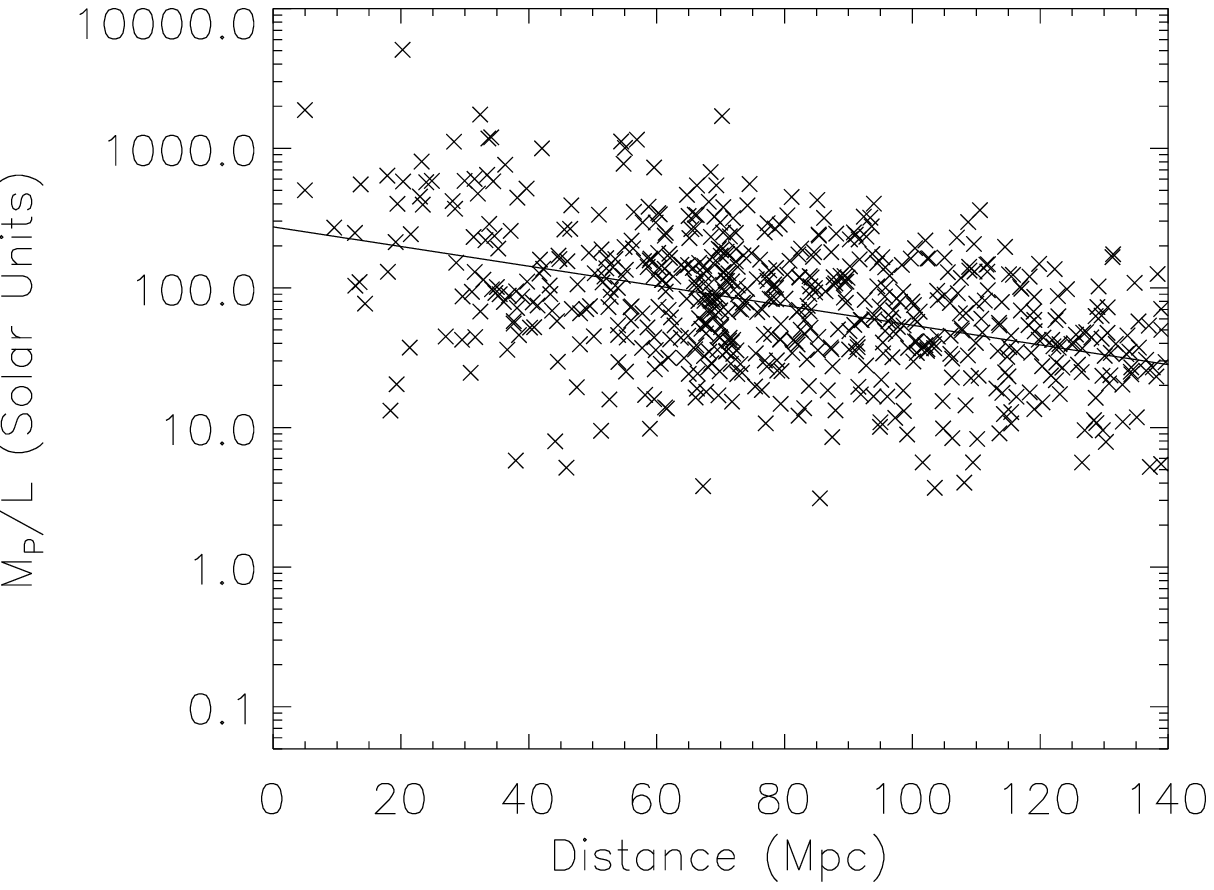}
\hspace{5mm}
\includegraphics[width=0.45\textwidth]{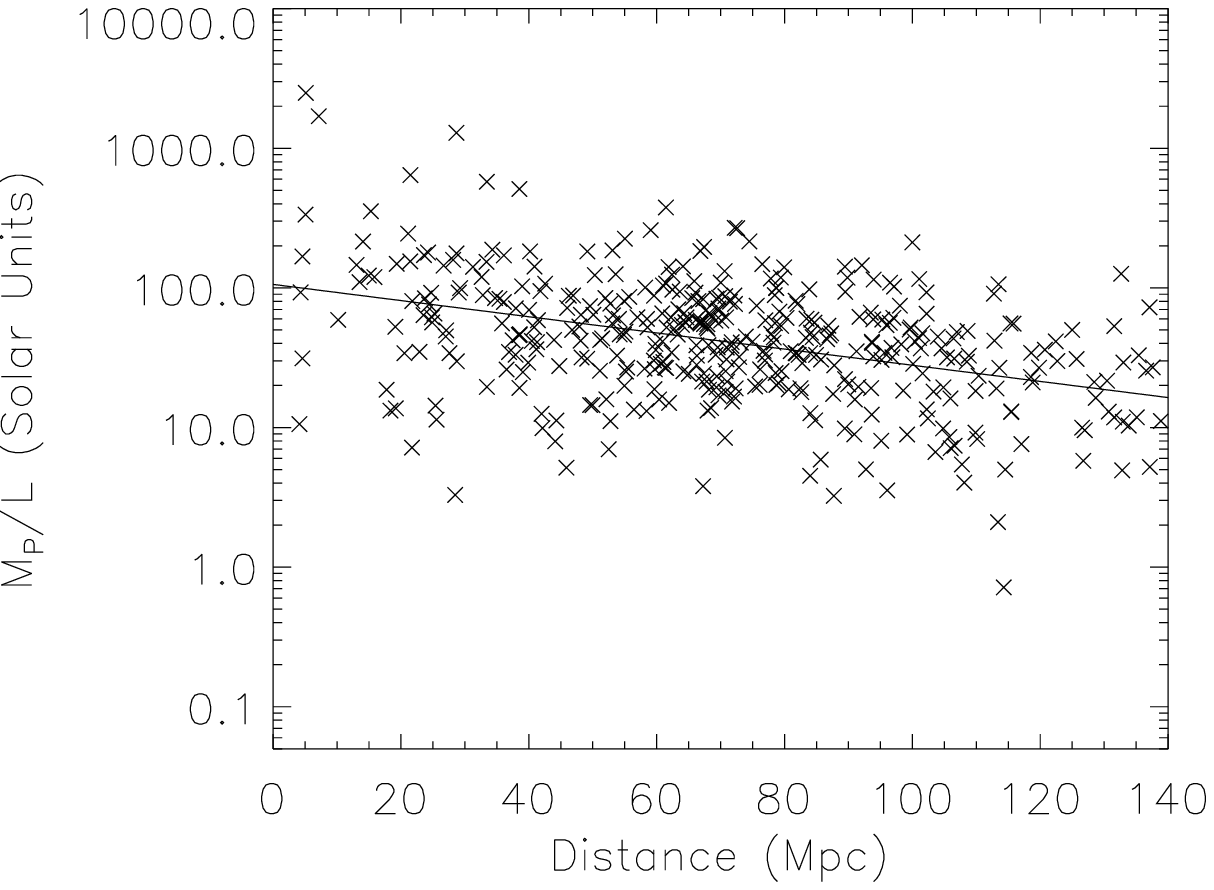}
\end{center}
\caption{\label{Fig:MassLight}Mass-to-light ratio of groups as a function of distance. The plots show the ratio of the projected mass estimate to the corrected K-band luminosity of the cluster.
The left panel shows the groups in the LDC catalog (\DCMax). The right panel shows the groups in the HDC catalog (\DCStd).
Only groups with 5 or more genuine members are included. The solid line shows a linear fit to the data as described in the text.}
\end{figure*}

Figure \ref{Fig:MassLight} shows the mass-to-light ratios (computed using the projected mass estimators) as a function of distance for both pairs of parameters.
%
%\placefigten
%\clearpage
%\clearpage
%
We fit a curve of the form
\begin{equation} \label{Eq:MLRatioFit}
\frac{M_P}{L_K} = 10^{(p D + q)}~\frac{\Msun}{\Lsun}
\end{equation}
to the data, where the best-fit parameters are given in Table~\ref{Tab:BestFitMLVD}.
%
%\clearpage
% Table 3
\begin{table}
\caption{\label{Tab:BestFitMLVD}Velocity-dispersion-- and $M/L$--distance relations}
\begin{center}
\begin{tabular}{ccc}
\tableline\tableline
Parameter                       & LDC Catalog         & HDC Catalog \\
\tableline
$\alpha$ ($10^{-3}$ Mpc$^{-1}$) &  $1.0 \pm 0.3$      &  $1.2 \pm 0.3$  \\
$\beta$                         &  $2.21 \pm 0.03$    &  $2.17 \pm 0.03$     \\
\tableline
$p$ ($10^{-3}$ Mpc$^{-1}$)      &  $-7.0 \pm 0.6$     &  $-5.8 \pm 0.6$  \\
$q$                             &  \Change{$2.43 \pm 0.05$}  &  \Change{$2.02 \pm 0.05$}  \\
\tableline
\end{tabular}
\end{center}
\tablecomments{Values correspond to the parameters in equations (\ref{Eq:VDFit}) and (\ref{Eq:MLRatioFit}) that minimize their respective $\chi^2$--statistic.}
\end{table}

%\clearpage%
The lower limit to the mass-to-light ratios computed as a function of distance remains approximately constant, while the upper limit decreases with distance, giving rise to the negative slope. Due to the nature of the flux-limited sample, at the largest distances we are not sensitive to (intrinsically) faint objects, therefore we are preferentially selecting the brightest groups. The scaling of the linking length, $D_L$, is designed to produce groups with a similar number of members at all distances. Since we are further correcting the luminosity of the groups to account for those galaxies to which the survey was not sensitive, we expect the mean luminosity of groups to increase with distance. As we have already shown that the velocity dispersions of the groups we find (and hence the estimated masses) are comparatively uncorrelated with distance, we would expect that we should miss those groups with high mass-to-light ratios at the largest distances, as indeed we observe in Figure \ref{Fig:MassLight}. To correct for this effect, one may introduce a scaling in the linking length in velocity space, $V_L$, with distance. This would increase the estimated group mass with distance, however such mass estimates would be based on groups containing many interlopers and thus not accurately represent the mass of the group. Such shortcomings of the percolation algorithm will be discussed in more detail in follow-up 
% In Prep Reference
work.\footnote{See Crook et al. (2006a), \textit{in preperation}.}

\subsubsection{Mass functions} \label{SubSection:MassFunctions}

\begin{figure*}
\begin{center}
\includegraphics[width=0.45\textwidth]{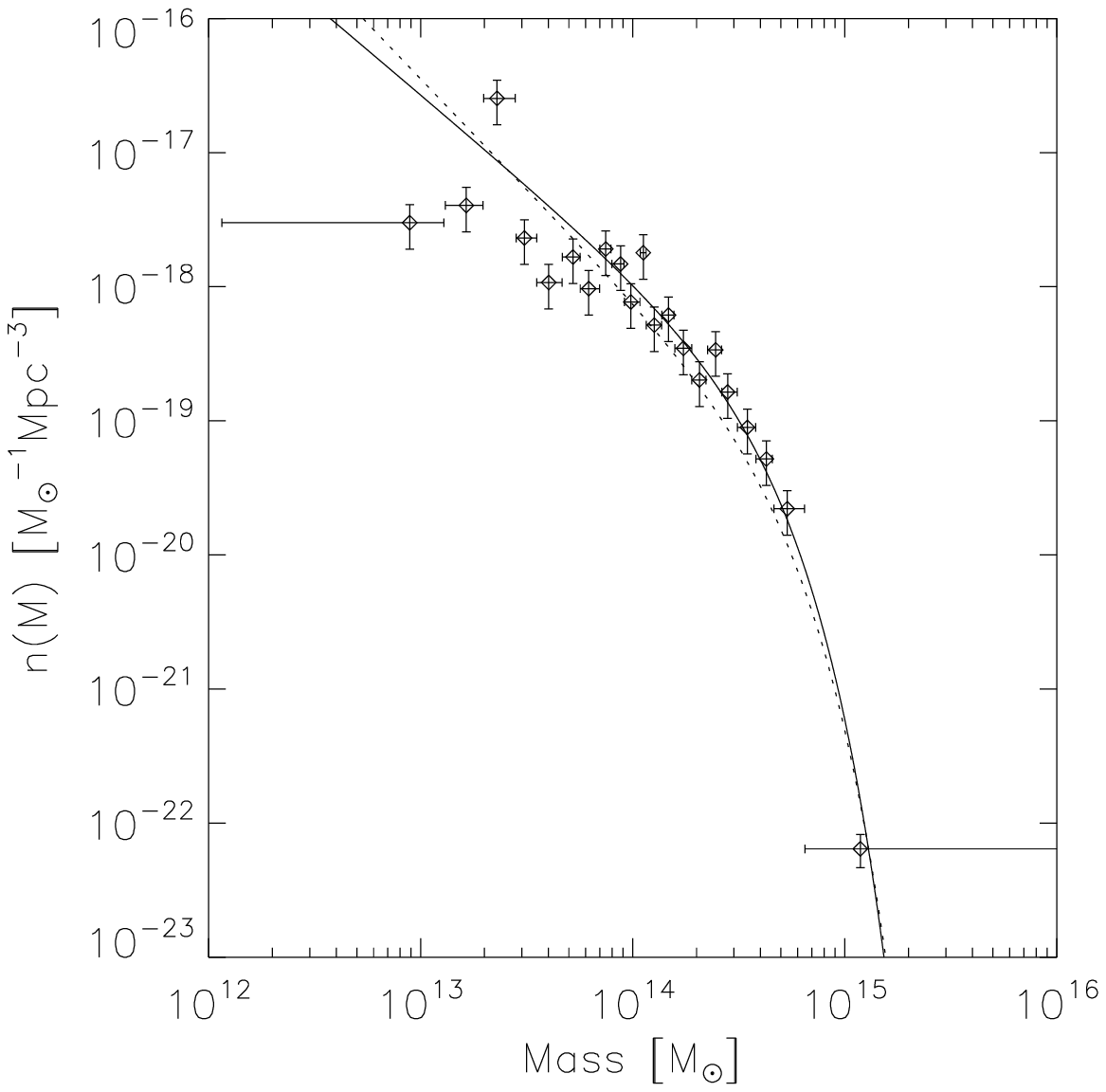}
\hspace{5mm}
\includegraphics[width=0.45\textwidth]{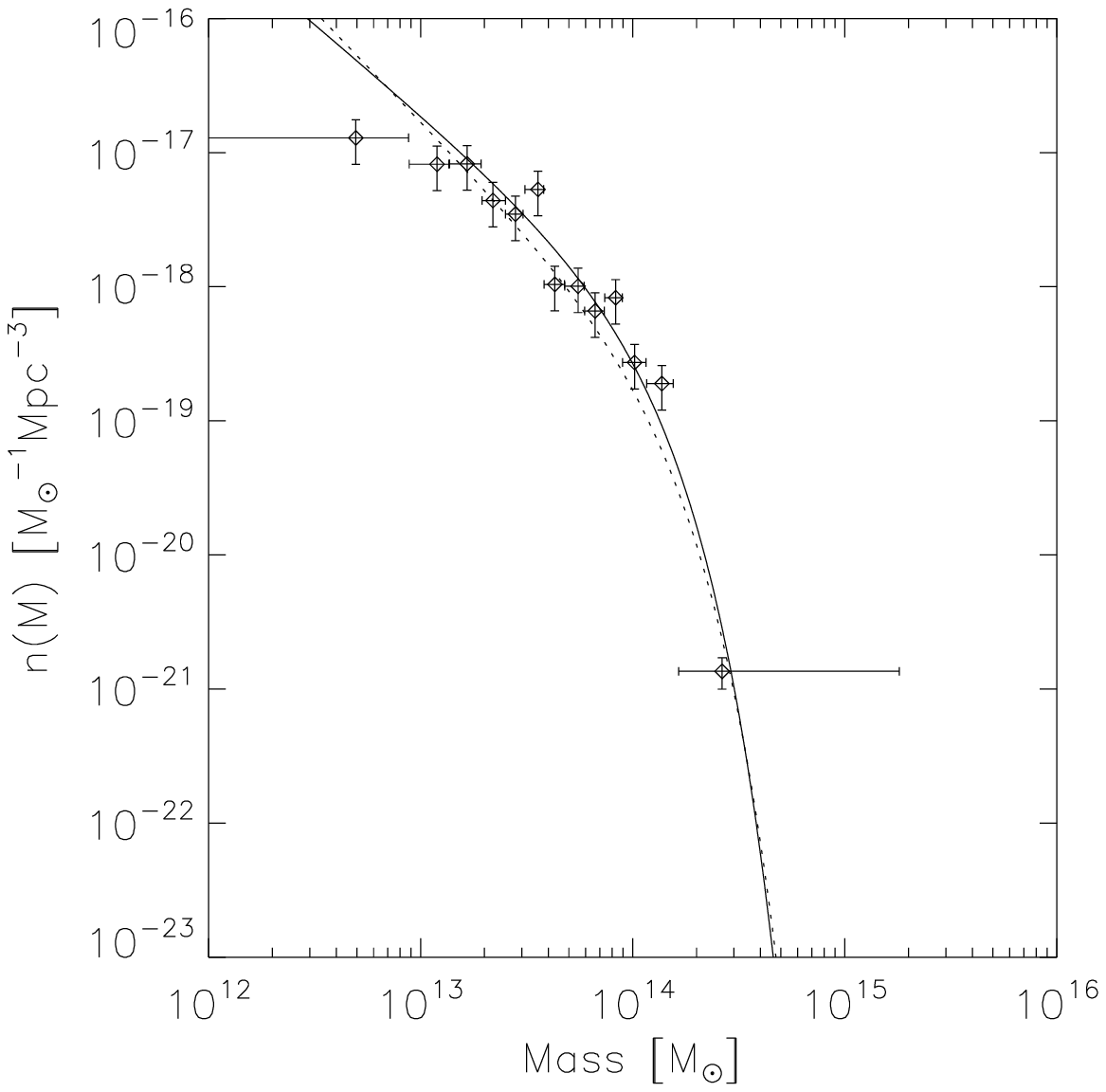}
\end{center}
\caption{\label{Fig:MassFunctions}Mass functions of the groups. 
The left panel shows the differential mass function of the groups in the LDC catalog (\DCMax); the right panel shows that of the groups in the HDC catalog (\DCStd).
The masses are estimated using the projected mass estimator. 2$\sigma$ error-bars are shown on the plots. Only groups at a distance of at least 10 Mpc containing at least 5 genuine members are included. The points have been fitted with a Press-Schechter (PS74) mass function (solid line) and a Sheth-Tormen (ST99) mass function (dotted line). The former fit produces the smaller $\chi^2$--statistic in both cases. \HAssumption
}
\end{figure*}

The large number of groups in the sample allows us to obtain an accurate estimate of the mass function for groups in the LDC and HDC catalogs. In this section we only consider groups with at least 5 genuine members at a distance of $10/h$ Mpc or greater.
We compute the mass function using the $1/V_\mathrm{max}$ procedure \citep[e.g.][]{Martinez:2002}, whereby each group is weighted by the inverse of the maximum comoving volume, $V_\mathrm{max}(L_i)$, in which the group remains observable given the flux limit of the survey. $L_i$ is the luminosity of the fifth brightest member of the group. The differential mass function can be then computing as
\begin{equation}
n(M) = \displaystyle\sum_{|M_i - M| \le \Delta M} [ V_\mathrm{max}(L_i) ]^{-1}
\end{equation}
where $M_i$ are the group masses and $\Delta M$ is the (variable) bin width.
The results are shown in Figure \ref{Fig:MassFunctions}.
%
%\placefigeleven
%\clearpage
%\clearpage
%

We consider analytical differential mass functions of the form suggested by \citet{Sheth:1999} \citep[also see][]{Jenkins:2001},
\begin{eqnarray} \label{Eq:MassFuncForm}
n(M) =~& A\frac{\gamma \bar{\rho}}{M^2} \sqrt{\frac{a}{\pi}}  \left( \frac{M}{M_0} \right)^{\gamma/2} \left[1+\frac{1}{(2a)^p} \left( \frac{M}{M_0} \right)^{-\gamma p} \right] \nonumber \\
  & \times \exp \left[-\left( \frac{M}{M_0} \right)^\gamma \right]
\end{eqnarray}
where $\gamma=1+(\tilde{n}/3)$ and we set $\tilde{n}=1$. The choice of parameters $a=1$, $p=0$, $A=0.5$ corresponds to the analytical prediction of \citet{Press:1974} (hereafter, PS74). \citet{Sheth:1999} suggest the alternative choice parameters $A=0.3222$, $a=0.707$, $p=0.3$ (hereafter, ST99), which provide good agreement with a subset of $N$-body simulations analyzed by \citet{Jenkins:2001}.

We fit functions of the form of equation (\ref{Eq:MassFuncForm}) to groups in the LDC and HDC catalogs using the parameter choices of both PS74 and ST99 (see Figure \ref{Fig:MassFunctions} and Table \ref{Tab:MassFuncBestFit}). 
%
%\clearpage
% Table 4
\begin{table}
\caption{\label{Tab:MassFuncBestFit}Best-fit values for Press-Schechter and Sheth-Tormen mass functions}
\begin{center}
\begin{tabular}{ccc}
\tableline\tableline
Parameter                                & LDC Catalog        & HDC Catalog \\
\tableline
\multicolumn{3}{c}{PS74 form\tablenotemark{a}} \\
\tableline
$\bar{\rho}$ ($10^{10}$ \Msun Mpc$^{-3}$)  & $2.8 \pm 1.4$      & $0.9 \pm 0.6$ \\
$\log ( M_0 / \Msun )$                   & $14.50 \pm 0.13$   & $13.93 \pm 0.18$ \\
\tableline
\multicolumn{3}{c}{ST99 form\tablenotemark{b}} \\
\tableline
$\bar{\rho}$ ($10^{10}$ \Msun Mpc$^{-3}$)  & $3.6 \pm 1.9$      & $1.3 \pm 0.8$ \\
$\log ( M_0 / \Msun )$                   & $14.43 \pm 0.16$   & $13.87 \pm 0.21$ \\
\tableline
\end{tabular}
\end{center}
\tablecomments{Values correspond to the parameters in equation (\ref{Eq:MassFuncForm}) that minimize the $\chi^2$--statistic.}
\tablenotetext{a}{\citet{Press:1974}}
\tablenotetext{b}{\citet{Sheth:1999}}
\end{table}

%\clearpage
%
The analytical descriptions are both good approximations to the data; similar conclusions were also drawn by \citet{Martinez:2002}, however we find the fit to the PS74 form produces a slightly smaller $\chi^2$--statistic in both cases.

We compare the ratio of the best-fit values for $M_0$ in the PS74 form of the mass function to the value predicted using the simple arguments of PS74.
$M_0$ scales with the minimum density contrast according to
\begin{equation}
M_0 \propto \left( \frac{\delta \rho}{\rho} \right)^{-2/\gamma}
\end{equation}
and thus we expect the ratio of the determined values of $M_0$ to be given by
\begin{eqnarray*}
\log \frac{M_0^\mathrm{(LDC)}}{M_0^\mathrm{(HDC)}}~ & = \log \left[ \frac{ (\delta \rho / \rho )_\mathrm{LDC} }{ ( \delta \rho / \rho)_\mathrm{HDC} } \right]^{-2/\gamma} \\
 & = -\frac{2}{\gamma} \log \left( \frac{12}{80} \right) = 1.24
\end{eqnarray*}
The obtained values of $M_0$ give log-ratio of $0.57 \pm 0.22$, providing agreement only at the $3\sigma$-level with Press-Schechter theory. The two methods cannot be expected to be within perfect agreement as the group-identification algorithm will find different groups as $V_0$ is varied but $D_0$ is held constant. The computed estimates of the group masses will therefore vary, while the ratio predicted by the Press-Schechter treatment (which is sensitive only to the change in density contrast, thus $D_0$) does not.

\subsubsection{Comparison with other group catalogs} \label{SubSection:OtherCatalogs}

To verify the validity of the produced group catalogs, we compare the 2MASS group catalogs with the UZC-SSRS2 group catalog \citep{Ramella:2002} and the CfAN group catalog \citep{Ramella:1997}. The former is constructed from partial versions of the Updated Zwicky Catalog \citep[UZC;][]{Falco:1999} and the Southern Sky Redshift Survey \citep[SSRS2;][]{daCosta:1998} and covers 37\% of the sky; the latter covers 10\%. The selection criteria for the three catalogs result in completeness limits at different velocities. To compute the completeness in redshift-space, we use the same technique as \S\ref{SubSection:VelocityLimit} above, however in this case, we determine the velocity corresponding to the peak in the derivative of equation (\ref{Eq:VelLimitFit}). When comparing the 2MASS group catalog with the other two, we cut both catalogs at the smallest of the two corresponding velocity limits (2MRS at 5697~\kms~in the LDC catalog and 5350~\kms~in the HDC catalog, UZC at 7115 \kms, CfAN at 9390 \kms). 

For each of the groups present in the UZC-SSRS2 and CfAN catalogs, we search the 2MRS group catalogs for a group within a radius set by twice the sum of the virial radii of the group in 2MRS and the group in the comparison catalog, with mean velocities that differ by less than 30\%. We find that 86\% of the groups in UZC-SSRS2 and 76\% of the groups in CfAN are present in the 2MRS LDC group catalog, whereas 78\% of the groups in UZC-SSRS2 and 69\% of the groups in CfAN are present in the 2MRS HDC group catalog. There is reasonable agreement in both cases given the differing selection criteria employed in the three samples.

\begin{figure*}
\begin{center}
\includegraphics[width=0.45\textwidth]{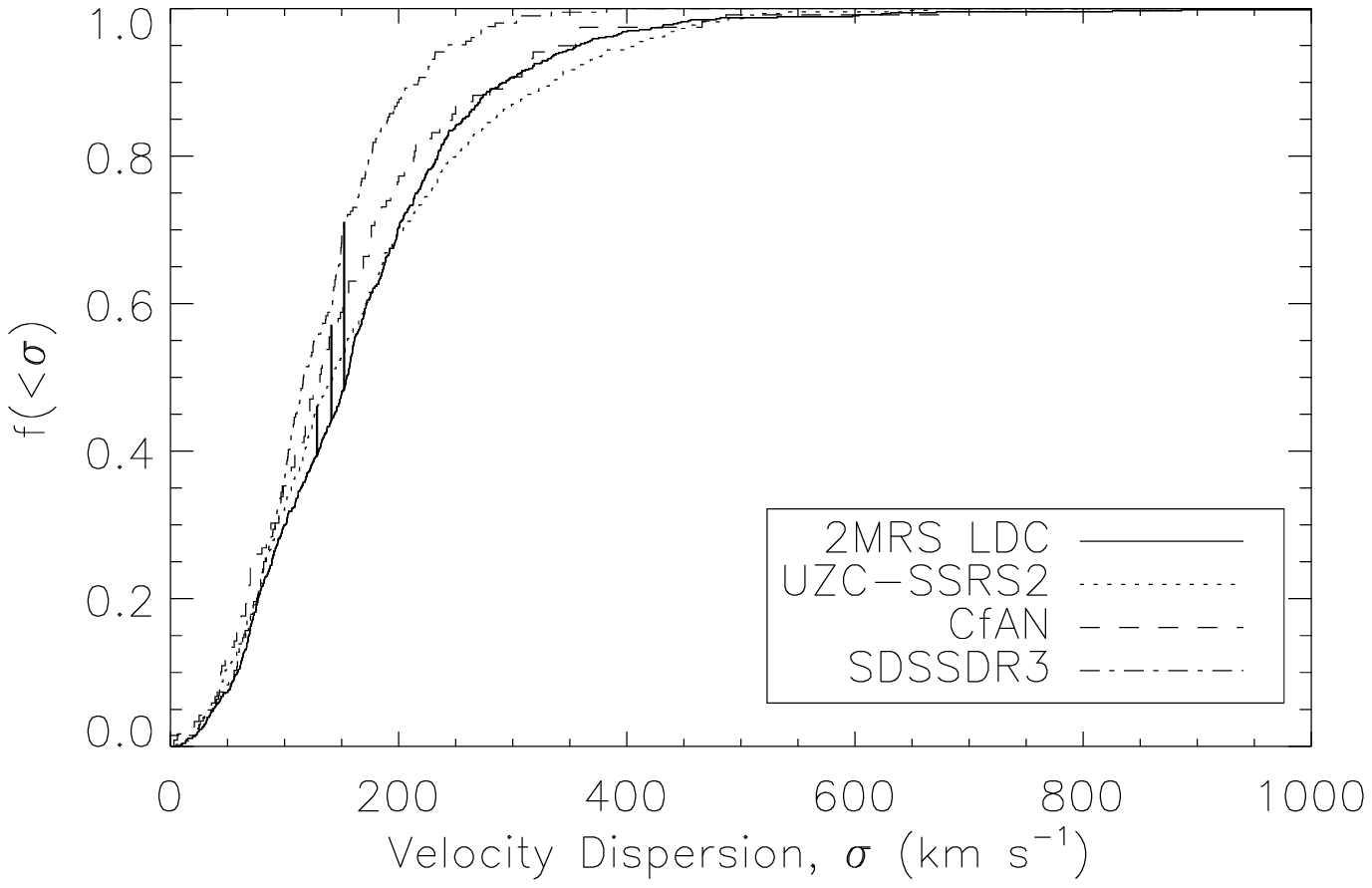}
\hspace{5mm}
\includegraphics[width=0.45\textwidth]{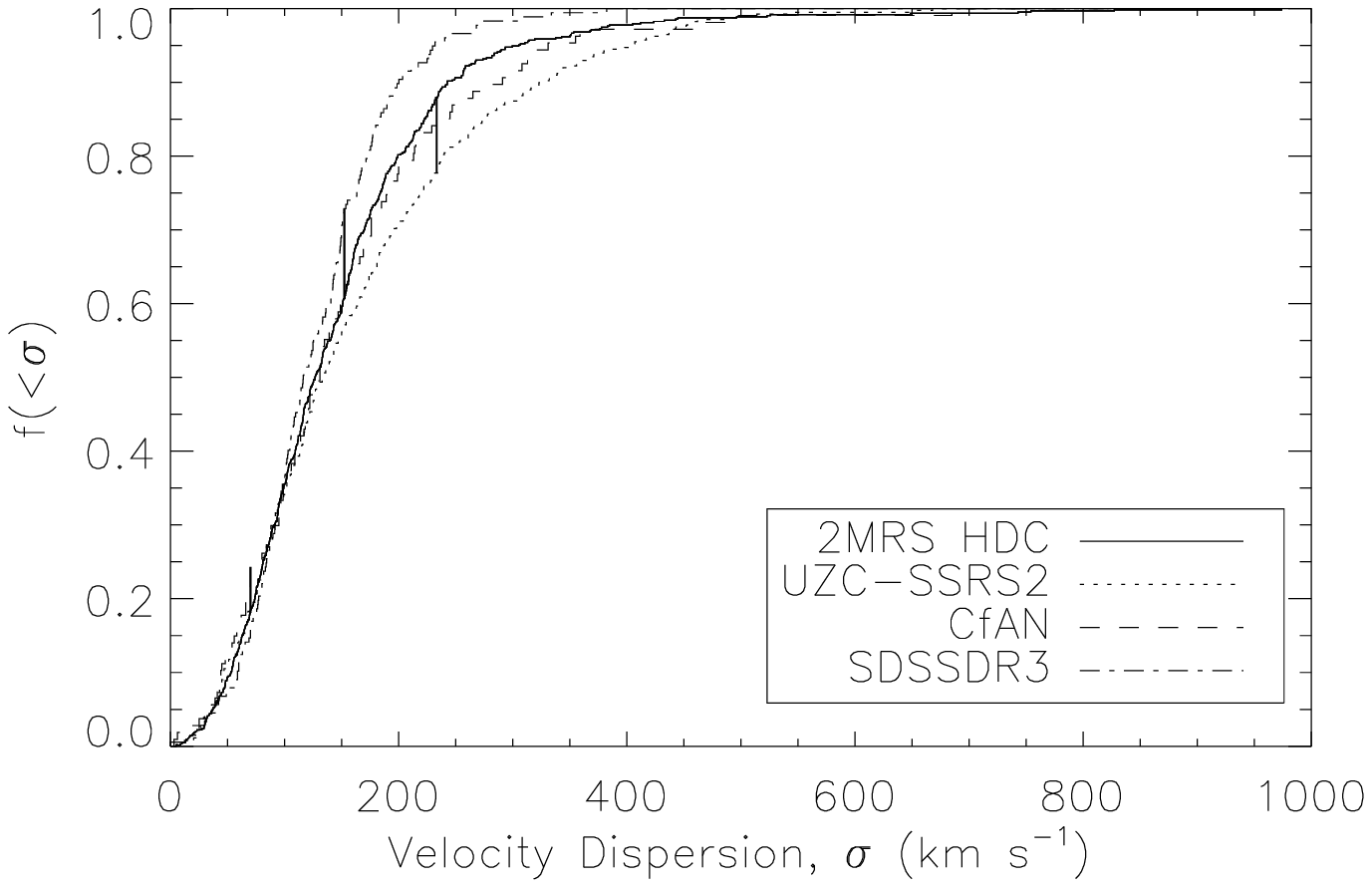}
\end{center}
\caption{\label{Fig:KSVelDisp}Cumulative fraction of velocity dispersions of groups. The velocity dispersions in the UZC-SSRS2, CfAN and SDSSDR3 group catalogs catalogs are compared with the 2MRS LDC catalog (left panel) and the 2MRS HDC catalog (right panel). The vertical lines indicate the corresponding $D$-statistic used in the K-S test.
}
\end{figure*}

We also compare the distributions of velocity dispersion using the two-sided Kolmogorov-Smirnov (K-S) test \citep{Chakravarti:1967}. The cumulative fractions of the LDC and HDC catalogs are compared with the cumulative fractions of the UZC-SSRS2 and CfAN catalogs (see Figure \ref{Fig:KSVelDisp}). 
For comparison, we also show the velocity dispersions of groups in 
Sloan Digital Sky Survey Data Release 3 \citep[SDSSDR3,][]{Merchan:2005}.
The results of the test are shown in Table \ref{Tab:VelDispKSResults}.
%
%\placefigtwelve
%\clearpage
%\clearpage
% Table 5
\begin{table*}
\caption{\label{Tab:VelDispKSResults}Comparison between 2MRS, UZC-SSRS2, CfAN and SDSSDR3 groups}
\begin{center}
\begin{tabular}{ccc}
\tableline\tableline
Property      & LDC Catalog    & HDC Catalog \\
\tableline
No. of 2MRS Groups  & 736 & 630 \\
\tableline
\multicolumn{3}{c}{Comparison with UZC-SSRS2} \\
\tableline
No. of UZC-SSRS2 Groups & 444 & 399 \\
$D$-statistic\tablenotemark{a} &  0.070 & 0.102 \\
$P$-value\tablenotemark{b}     &  0.128 & 0.011 \\
\tableline
\multicolumn{3}{c}{Comparison with CfAN} \\
\tableline
No. of CfAN Groups & 119 & 107 \\
$D$-statistic\tablenotemark{a} &  0.128 & 0.062 \\
$P$-value\tablenotemark{b}     &  0.062 & 0.862 \\
\tableline
\multicolumn{3}{c}{Comparison with SDSSDR3} \\
\tableline
No. of SDSSDR3 Groups & 204 & 177 \\
$D$-statistic\tablenotemark{a} &  0.227     & 0.122 \\
$P$-value\tablenotemark{b}     &  $10^{-7}$ & 0.029 \\
\tableline
\end{tabular}
\end{center}
\tablecomments{The table presents the results of two-sided Kolmogorov-Smirnov test on the comparison between the distributions of velocity dispersions of the groups in 2MRS catalog with the UZC-SSRS2, CfAN and SDSSDR3 catalogs.}
\tablenotetext{a}{$D$-statistic used in computation of $P$-value in Kolmogorov-Smirnov test.}
\tablenotetext{b}{$P$-value represents the probability that such a difference would be observed under the assumption that the two samples were drawn from the same parent distribution. We consider values of $P < 0.05$ to indicate that the two samples were drawn from significantly different parent distributions.}
\end{table*}

%\clearpage
%
The velocity dispersions in both the LDC and HDC catalogs are consistent with the CfAN sample. However only the LDC catalog produces groups with velocity dispersions consistent with those in the UZC-SSRS2 catalog, although the discrepancy with the HDC catalog is small. Any discrepancy with the CfAN catalog is not appreciable due to the smaller number of groups in the compared sample of CfAN galaxies.\footnote{This is accounted for in the Kolmogorov-Smirnov test.} 
Neither the LDC or HDC catalogs are consistent with the SDSSDR3 group catalog. The discrepancy is expected due to the difference in the algorithms used to identify groups.

The difference in the distributions of velocity dispersion is due in part to the different scalings of the velocity-linking parameter. When scaling the linking length, the more distant groups are likely to have higher velocity dispersion. These groups may not be present in a group catalog derived by setting the parameter to a constant.
The velocity dispersions of groups in the USZ-SSRS2 and CfAN catalogs tail off at $\sim$400~\kms. The LDC catalog is in reasonable agreement with this since the velocity linking parameter is $\sim$400~\kms. The HDC catalog, however, contains a smaller fraction of groups with velocity dispersions between 350--400~\kms~than are present in the other catalogs. This is consistent with the above hypothesis since, in this case, the velocity linking parameter is set to 350~\kms~in this case. Without knowledge of the complete phase-space positions of the galaxies, we cannot determine whether the majority of groups with velocity dispersions greater than 350~\kms~in the LDC, UZC-SSRS2 and CfAN catalogs are bound or not.

\begin{figure*}
\begin{center}
\includegraphics[width=0.9\textwidth]{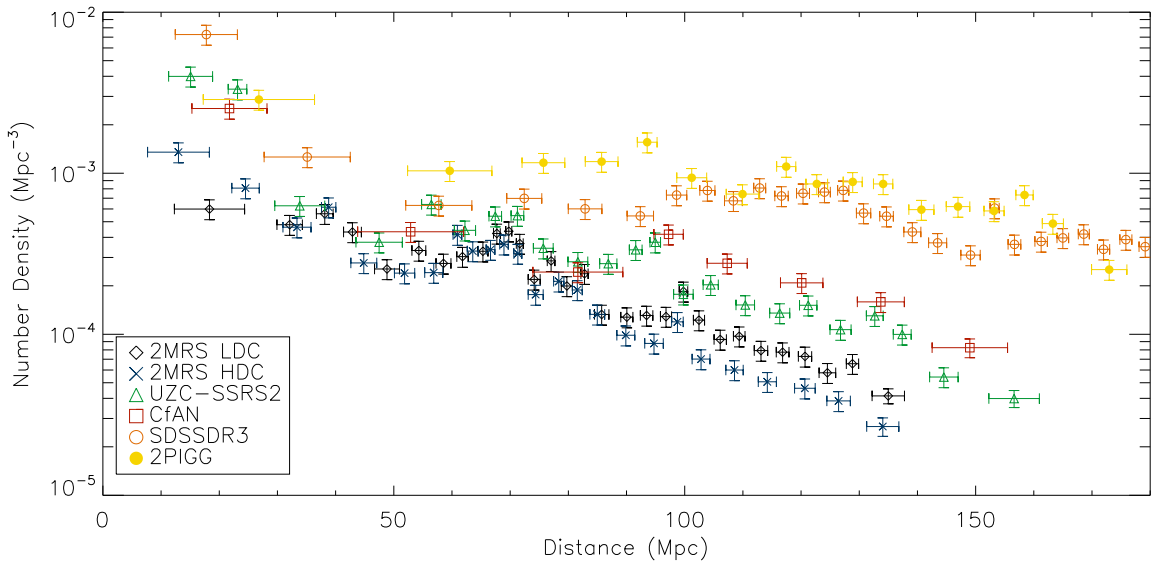}
\end{center}
\caption{\label{Fig:NumberDensity}Number density of groups as a function of distance. We show the number density for the LDC (diamond) and HDC (cross) catalogs, as well as the UZC-SSRS2 (triangle), CfAN (square), SDSSDR3 (open circle) and 2PIGG (filled circle) catalogs. (The symbols are colored black, blue, green, red, orange and yellow, respectively, in the electronic edition.) \HAssumption
}
\end{figure*}

We further examine the number density of groups as a function of redshift (see Figure \ref{Fig:NumberDensity}). For comparison, we include the number density of groups in SDSSDR3 and the 2dFGRS Percolation-Inferred Galaxy Group (2PIGG) catalog \citep{Eke:2004}, although these surveys have very different selection biases and fainter flux limits, which result in a higher number density of groups than that detected in 2MRS.
%
%\placefigthirteen
%\clearpage
%\clearpage
%
The 2MRS group catalogs contain an approximately constant number density of groups to $\sim$80 Mpc, before following a power-law decay. The LDC catalog has slightly more uniform coverage out to this distance than the HDC catalog. The observed shape of the function is consistent between the LDC, HDC, UZC-SSRS2 and CfAN catalogs, and the values only diverge at $\sim$90 Mpc where the linking length becomes large enough that the selection biases of the parent surveys will have a significant effect on the identified groups. The number density of groups in the SDSSDR3 and 2PIGG catalogs is approximately constant over the entire range of distances considered due to the lower flux limits of the surveys. We note that the HDC catalog contains fewer groups than the LDC catalog beyond $\sim$30 Mpc, as expected. The observed peak at $\sim$70 Mpc corresponds to the location of the great attractor, and is enhanced by the distance corrections discussed in \S\ref{SubSection:Flowfield}.

\section{Summary} \label{Section:Summary}
We have presented two catalogs of groups in the 2MASS Redshift Survey, identified using a variable-linking-length percolation algorithm (HG82). We discussed the effect of the variation of the input parameters, $D_0$ and $V_0$, on the number of groups obtained. 
As demonstrated in \S\ref{SubSection:Reliability} above, we see that the correct choice of parameters depends on the purpose of the catalog, and full phase-space information of each galaxy is required to understand the most suitable choice of parameters to find virialized
% In prep reference
groups.\footnote{See follow-up work, Crook et al. \Change{(2007a)}, \textit{In preparation}.}

We justify the choice of two pairs of parameters: \ParamMax, corresponding to a density contrast $\delta \rho / \rho$ = \DensityContrastMax~(LDC catalog), and \ParamStd, corresponding to a density constrast $\delta \rho / \rho = 80$ (HDC catalog). We show that the latter choice of parameters identifies the largest nearby clusters individually, while many of these groups are merged with the former parameter choice.

We compute virial and projected mass estimates for the clusters under the assumptions that the identified groups have spherical symmetry and that the light traces the distribution of the mass.
We find that the projected mass estimates give mass functions in agreement at the 3$\sigma$-level with Press-Schechter theory, although given the dependence of the algorithm on $V_0$, 1$\sigma$-agreement is not expected.

We calculate corrected K-band luminosities for each cluster, and use these to estimate the mass-to-light ratios and corresponding values of $\Omega_m$. The values predicted using the virial mass estimator in both the LDC and HDC catalogs are significantly smaller than the 3-year WMAP result \citep{Spergel:2006} of $\Omega_m = 0.238^{+0.013}_{-0.024}$ motivating the use of the projected mass estimator over the virial mass estimator in subsequent analysis. The projected mass estimates of groups in the LDC catalog produce to a value of $\Omega_m = 0.229_{-0.012}^{+0.016}$, which agrees with the WMAP result at the 1$\sigma$-level. The HDC catalog significantly under-predicts the WMAP value, suggesting that by only including groups with density contrasts $\delta \rho / \rho \ge 80$ we are underestimating the total mass in groups.

The distribution of velocity dispersions of groups in the 2MRS LDC catalog is in agreement with the groups in the UZC-SSRS2 catalog \citep{Ramella:2002} as well as CfAN group catalog. The 2MRS HDC catalog velocity dispersions are in agreement with the groups of the CfAN catalog, however we find that there is a statistically significant difference between the distributions of velocity dispersions in the 2MRS HDC and UZC-SSRS2 catalogs.

The group catalogs presented in this paper provide an estimate of cluster locations and masses without the necessity to assume an intrinsic mass-to-light ratio. We use and discuss the results of the clustering analysis in 
% In Prep Reference
follow-up work,\footnote{See Crook et al. \Change{(2007b)}, \textit{In preparation}.}
including the development of a flow-field model to estimate the discrepency between the expected flow of the local group and the observed dipole in the cosmic microwave background \citep{Bennett:1996,Bennett:2003}. These group catalogs form the basis for a map of baryonic density enhancements in the nearby Universe, which can be compared with flow-field maps developed using peculiar-velocity surveys in order to infer the presence and location of dark matter that is not correlated with luminous matter.

\acknowledgments
We thank the observers at the Fred Lawrence Whipple Observatory's 1.5m telescope, especially Perry Berlind and Mike Calkins, who obtained the majority of the spectra of galaxies in the northern hemisphere. We thank Susan Tokarz at the Harvard-Smithsonian Center for Astrophysics for reducing most of these spectra. We also thank Paul Schechter, Genevieve Monsees and Ben Cain for useful discussions\Change{, and Igor Karachentsev and Xiaohu Yang for their comments on this work.}
We thank those who have contributed to the 2MASS Redshift Survey (including Steve Schneider, Roc Cutri, Jeff Mader, Ted George, Mike Skrutskie, Chris Kochanek, Emilio Falco, and Mike Pahre) and the 6dF team (including Will Saunders, Quentin Parker, Matthew Colless, Ofer Lahav, Gary Mamon, John Lucey, Elaine Sadler, Fred Watson, Dominique Proust, and Ken-ichi Wakamatsu).
This publication makes use of data products from the Two Micron All Sky Survey, which is a joint project of the University of Massachusetts and the Infrared Processing and Analysis Center/California Institute of Technology, funded by the National Aeronautics and Space Administration and the National Science Foundation. This publication also makes use of data products from the 6dF Survey of the AATO.
This research has also made use of source code available at \mbox{http://www.dfanning.com/}. This work is supported by NSF grant AST 0406906.

\appendix

\section{Group catalogs}
%\clearpage
% Table A1
\clearpage
\begin{landscape}
\begin{deluxetable}{cccccccccccccc}
%\rotate
\tabletypesize{\tiny}
\tablecolumns{14}
\tablewidth{0pt}
\tablecaption{\label{Tab:GroupCatalogMax}Low-density-contrast (LDC) catalog of groups in the 2MASS Redshift Survey}
\tablehead{
\colhead{\# } & \colhead{RA } & \colhead{Dec } & \colhead{Members\tablenotemark{a} } & \colhead{Distance\tablenotemark{b} } & \colhead{${V_G}$\tablenotemark{c} } & \colhead{${\sigma_P}$\tablenotemark{d} } & \colhead{${R_\mathrm{P}}$\tablenotemark{e} } & \colhead{$\log \left[ \frac{M_V}{M_\odot} \right]$\tablenotemark{f} } & \colhead{$\log \left[ \frac{M_P}{M_\odot} \right]$\tablenotemark{g} } & \colhead{$\log \left[ \frac{M_P / L}{M_\odot / L_\odot} \right]$\tablenotemark{i} } & \colhead{$a$\tablenotemark{j} } & \colhead{$\eta$\tablenotemark{k} } & \colhead{$\phi$\tablenotemark{l}} \\
 & & & & \colhead{(Mpc)} & \colhead{(km/s)} & \colhead{(km/s)} & \colhead{(Mpc)} & & & & \colhead{($'$)} & & \colhead{($^\circ$)}
}
\startdata
    1 & 00$^h$00$^m$09.5$^s$ & $+$32$^\circ$44$'$28$''$ &       3 (0) &  136.62 & 10087 &   284.5 &  0.45 &  13.597 &  13.398 &   1.121 &     5 &      &     \\
    2 & 00$^h$00$^m$48.3$^s$ & $+$04$^\circ$05$'$18$''$ &       3 (0) &  119.62 &  8870 &   272.9 &  3.29 &  14.429 &  14.320 &   2.333 &    39 &      &     \\
    3 & 00$^h$00$^m$50.1$^s$ & $+$28$^\circ$17$'$00$''$ &       5 (0) &  120.31 &  8894 &   199.7 &  0.93 &  13.609 &  13.930 &   1.544 &    33 & 0.22 &  80 \\
    4 & 00$^h$02$^m$07.0$^s$ & $+$06$^\circ$57$'$55$''$ &       4 (0) &   71.53 &  5306 &    74.7 &  4.66 &  13.455 &  13.505 &   1.775 &   122 &      &     \\
    5 & 00$^h$02$^m$28.8$^s$ & $-$54$^\circ$29$'$49$''$ &       3 (0) &  132.62 &  9652 &   198.6 &  0.02 &  12.002 &  14.117 &   1.698 &    33 &      &     \\
    6 & 00$^h$05$^m$31.5$^s$ & $+$27$^\circ$29$'$37$''$ &       3 (0) &  102.61 &  7590 &    86.8 &  1.95 &  13.207 &  13.206 &   1.457 &    29 &      &     \\
    7 & 00$^h$05$^m$41.8$^s$ & $+$05$^\circ$09$'$11$''$ &       3 (0) &   71.94 &  5340 &    41.0 &  0.64 &  12.072 &  11.974 &   0.553 &    15 &      &     \\
    8 & 00$^h$06$^m$25.3$^s$ & $-$52$^\circ$12$'$28$''$ &       3 (0) &  138.95 & 10130 &   289.9 &  4.81 &  14.646 &  14.715 &   2.368 &    51 &      &     \\
    9 & 00$^h$06$^m$32.0$^s$ & $+$32$^\circ$25$'$02$''$ &      10 (0) &   66.21 &  4871 &   102.6 &  0.84 &  12.988 &  13.406 &   1.386 &    86 & 0.19 &  40 \\
   10 & 00$^h$06$^m$51.6$^s$ & $-$33$^\circ$40$'$17$''$ &       3 (0) &   93.51 &  6872 &    90.5 &  4.07 &  13.562 &  13.523 &   1.580 &    60 &      &     \\

\enddata
\tablenotetext{a}{No. of group members (including those generated from the population of the plane). The number derived from the galactic-plane population is contained in parentheses.}
\tablenotetext{b}{Mean (corrected) group distance.}
\tablenotetext{c}{Mean heliocentric group velocity.}
\tablenotetext{d}{Line-of-sight velocity dispersion.}
\tablenotetext{e}{Projected virial radius.}
\tablenotetext{f}{Log of the virial mass in solar units.}
\tablenotetext{g}{Log of the projected mass in solar units.}
\tablenotetext{i}{Log of the (projected) mass-to-light ratio in solar units.}
\tablenotetext{j}{Semi-major axis of the ellipse fit to the group at the 75$^\mathrm{th}$--percentile level (measured in arcminutes).}
\tablenotetext{k}{Axis-ratio of ellipse fit to the group members.}
\tablenotetext{l}{Position angle of semi-major axis of ellipse fit to the group members; measured from north toward east.}
\tablecomments{The complete version of this table is in the electronic edition of the Journal. The printed edition contains only a sample.
This catalog has been produced using parameters \ParamMax, corresponding to the density contrast \DCMax.
We assume \hvalue~where a value is required.}
\end{deluxetable}
\clearpage
\end{landscape}
% Table A2
\clearpage
\begin{landscape}
\begin{deluxetable}{ccccccccccccccc}
\tabletypesize{\tiny}
\tablecolumns{15}
%\rotate
\tablewidth{0pt}
\tablecaption{\label{Tab:GroupCatalogStd}High-density-contrast (HDC) catalog of groups in the 2MASS Redshift Survey}
\tablehead{
\colhead{\# } & \colhead{RA } & \colhead{Dec } & \colhead{Members\tablenotemark{a} } & \colhead{Distance\tablenotemark{b} } & \colhead{${V_G}$\tablenotemark{c} } & \colhead{${\sigma_P}$\tablenotemark{d} } & \colhead{${R_\mathrm{P}}$\tablenotemark{e} } & \colhead{$\log \left[ \frac{M_V}{M_\odot} \right]$\tablenotemark{f} } & \colhead{$\log \left[ \frac{M_P}{M_\odot} \right]$\tablenotemark{g} } & \colhead{$\log \left[ \frac{M_P / L}{M_\odot / L_\odot} \right]$\tablenotemark{i} } & \colhead{$a$\tablenotemark{j} } & \colhead{$\eta$\tablenotemark{k} } & \colhead{$\phi$\tablenotemark{l}} & Corresponding \\
 & & & & \colhead{(Mpc)} & \colhead{(km/s)} & \colhead{(km/s)} & \colhead{(Mpc)} & & & & \colhead{($'$)} & & \colhead{($^\circ$)} & Group \#\tablenotemark{n}
}
\startdata
    1 & 00$^h$00$^m$09.5$^s$ & $+$32$^\circ$44$'$28$''$ &       3 (0) &  136.62 & 10087 &   284.5 &  0.45 &  13.597 &  13.398 &   1.121 &     5 &      &     &     1 \\
    2 & 00$^h$00$^m$39.2$^s$ & $+$47$^\circ$05$'$02$''$ &      10 (0) &   70.88 &  5165 &   151.6 &  1.60 &  13.606 &  13.696 &   1.612 &    69 & 0.32 &  97 &    11 \\
    3 & 00$^h$02$^m$37.5$^s$ & $+$31$^\circ$20$'$47$''$ &       4 (0) &   66.16 &  4867 &    99.6 &  0.17 &  12.278 &  13.006 &   1.546 &    24 &      &     &     9 \\
    4 & 00$^h$05$^m$31.5$^s$ & $+$27$^\circ$29$'$37$''$ &       3 (0) &  102.61 &  7590 &    86.8 &  1.95 &  13.207 &  13.206 &   1.457 &    29 &      &     &     6 \\
    5 & 00$^h$05$^m$41.8$^s$ & $+$05$^\circ$09$'$11$''$ &       3 (0) &   71.94 &  5340 &    41.0 &  0.64 &  12.072 &  11.974 &   0.553 &    15 &      &     &     7 \\
    6 & 00$^h$09$^m$11.3$^s$ & $+$33$^\circ$07$'$36$''$ &       6 (0) &   66.25 &  4873 &   113.9 &  1.05 &  13.174 &  13.328 &   1.448 &    23 & 0.69 &  46 &     9 \\
    7 & 00$^h$10$^m$35.7$^s$ & $-$56$^\circ$59$'$21$''$ &       3 (0) &  132.31 &  9620 &   177.4 &  0.51 &  13.242 &  13.365 &   0.976 &     7 &      &     &    13 \\
    8 & 00$^h$12$^m$00.7$^s$ & $+$16$^\circ$09$'$37$''$ &       3 (0) &   13.51 &   912 &   101.3 &  1.18 &  13.123 &  13.216 &   2.349 &   174 &      &     &    16 \\
    9 & 00$^h$13$^m$05.5$^s$ & $+$30$^\circ$57$'$12$''$ &       3 (0) &   65.04 &  4792 &    61.3 &  0.56 &  12.360 &  12.424 &   0.929 &    12 &      &     &    15 \\
   10 & 00$^h$13$^m$18.8$^s$ & $+$22$^\circ$17$'$48$''$ &       3 (0) &   81.42 &  6034 &   121.2 &  0.58 &  12.973 &  13.454 &   1.839 &    31 &      &     &    14 \\
\enddata
\tablecomments{The complete version of this table is in the electronic edition of the Journal.  The printed edition contains only a sample.
This catalog has been produced using parameters \ParamStd, corresponding to the density contrast \DCStd.
We assume \hvalue~where a value is required.}
\tablenotetext{a}{No. of group members (including those generated from the population of the plane). The number derived from the galactic-plane population is contained in parentheses.}
\tablenotetext{b}{Mean (corrected) group distance.}
\tablenotetext{c}{Mean heliocentric group velocity.}
\tablenotetext{d}{Line-of-sight velocity dispersion.}
\tablenotetext{e}{Projected virial radius.}
\tablenotetext{f}{Log of the virial mass in solar units.}
\tablenotetext{g}{Log of the projected mass in solar units.}
\tablenotetext{i}{Log of the (projected) mass-to-light ratio in solar units.}
\tablenotetext{j}{Semi-major axis of the ellipse fit to the group at the 75$^\mathrm{th}$--percentile level (measured in arcminutes).}
\tablenotetext{k}{Axis-ratio of ellipse fit to the group members.}
\tablenotetext{l}{Position angle of semi-major axis of ellipse fit to the group members; measured from north toward east.}
\tablenotetext{n}{Group number from LDC catalog that encompasses all members of this group.}
\end{deluxetable}
\clearpage
\end{landscape}
% Table A3
\clearpage
\begin{landscape}
\begin{deluxetable}{ccccccccccccc}
\tabletypesize{\tiny}
\tablecolumns{13}
%\rotate
\tablewidth{0pt}
\tablecaption{\label{Tab:GroupCatalogBigMax}Groups in the LDC catalog with 50 or more members}
\tablehead{
\colhead{\#} & \colhead{RA} & \colhead{Dec} & \colhead{Members\tablenotemark{a}} & \colhead{Distance\tablenotemark{b}} & \colhead{${V_G}$\tablenotemark{c}} & \colhead{${\sigma_P}$\tablenotemark{d}} & \colhead{${R_\mathrm{P}}$\tablenotemark{e}} & \colhead{$\log \left[ \frac{M_V}{M_\odot} \right]$\tablenotemark{f}} & \colhead{$\log \left[ \frac{M_P}{M_\odot} \right]$\tablenotemark{g}} & \colhead{$\log \left[ \frac{M_P / L}{M_\odot / L_\odot} \right]$\tablenotemark{i}} & \colhead{Identified} \\
 & & & & \colhead{(Mpc)} & \colhead{(km/s)} & \colhead{(km/s)} & \colhead{(Mpc)} & & & & \colhead{With\tablenotemark{j}}
}
\startdata
  852 & 12$^h$19$^m$12.0$^s$ & $+$16$^\circ$56$'$52$''$ &     810 (2) &   19.46 &  1314 &   671.4 & 10.42 &  15.712 &  16.102 &   2.601 & Virgo + NGC3607 Cluster \\
  391 & 05$^h$29$^m$43.7$^s$ & $-$42$^\circ$50$'$28$''$ &    302 (22) &   20.25 &  1493 &   597.8 &  8.07 &  15.500 &  16.728 &   3.704 & Eridanus + Fornax I + NGC2280 Cluster + Dorado \\
  229 & 03$^h$08$^m$45.7$^s$ & $+$41$^\circ$47$'$55$''$ &     301 (0) &   72.39 &  5322 &  1027.5 &  4.95 &  15.758 &  15.982 &   2.394 & Perseus-Pisces (A426, A347) \\
  712 & 10$^h$24$^m$57.5$^s$ & $-$28$^\circ$29$'$04$''$ &     241 (0) &   46.47 &  3283 &   623.4 &  5.96 &  15.405 &  15.681 &   2.422 & Hydra (A1060, A1060) \\
 1117 & 16$^h$10$^m$44.6$^s$ & $-$59$^\circ$50$'$30$''$ &    217 (42) &   68.66 &  4815 &   825.7 &  4.35 &  15.512 &  15.723 &   2.221 & Norma (A3627) \\
  881 & 12$^h$51$^m$27.6$^s$ & $-$42$^\circ$11$'$34$''$ &     202 (0) &   52.79 &  3301 &   631.3 &  4.65 &  15.308 &  15.347 &   1.936 & Centaurus (A3526) \\
 1259 & 18$^h$57$^m$14.6$^s$ & $-$62$^\circ$19$'$13$''$ &     103 (0) &   62.79 &  4415 &   348.3 &  3.70 &  14.691 &  14.871 &   1.731 &  \\
   79 & 01$^h$15$^m$17.5$^s$ & $+$32$^\circ$47$'$01$''$ &      98 (0) &   67.60 &  5006 &   431.7 &  3.35 &  14.836 &  15.071 &   1.933 &  \\
  280 & 04$^h$03$^m$29.2$^s$ & $+$51$^\circ$34$'$16$''$ &     86 (78) &   71.02 &  5182 &   532.0 &  6.10 &  15.277 &  15.303 &   2.243 &  \\
  956 & 13$^h$39$^m$37.8$^s$ & $-$31$^\circ$05$'$29$''$ &      84 (0) &   58.92 &  4322 &   481.8 &  3.25 &  14.917 &  15.204 &   2.197 & Centaurus (A3574) \\
  890 & 12$^h$58$^m$53.8$^s$ & $+$27$^\circ$53$'$23$''$ &      84 (0) &  100.24 &  6842 &   648.0 &  3.31 &  15.183 &  15.271 &   1.876 & Coma (A1656) \\
  883 & 12$^h$54$^m$53.4$^s$ & $-$11$^\circ$16$'$36$''$ &      82 (0) &   62.08 &  4258 &   444.7 &  3.37 &  14.863 &  15.186 &   2.215 &  \\
  328 & 04$^h$41$^m$15.9$^s$ & $-$05$^\circ$08$'$37$''$ &      77 (0) &   61.24 &  4510 &   375.3 &  4.12 &  14.803 &  15.040 &   2.136 &  \\
 1454 & 22$^h$28$^m$34.8$^s$ & $+$35$^\circ$40$'$15$''$ &      76 (0) &   84.62 &  6143 &   532.0 &  4.23 &  15.118 &  15.445 &   2.255 &  \\
  127 & 01$^h$57$^m$18.5$^s$ & $+$34$^\circ$19$'$08$''$ &      71 (0) &   66.22 &  4905 &   388.1 &  3.05 &  14.702 &  14.966 &   2.003 & Perseus-Pisces (A262) \\
  811 & 11$^h$45$^m$24.1$^s$ & $+$20$^\circ$08$'$13$''$ &      56 (0) &   95.74 &  6548 &   550.8 &  1.83 &  14.785 &  14.825 &   1.691 & Coma (A1367) \\
  292 & 04$^h$14$^m$23.7$^s$ & $+$36$^\circ$58$'$31$''$ &      51 (0) &   81.93 &  6017 &   286.1 &  3.49 &  14.496 &  14.810 &   1.826 &  \\
\enddata
\tablenotetext{a}{No. of group members (including those generated from the population of the plane). The number derived from the galactic-plane population is contained in parentheses.}
\tablenotetext{b}{Mean (corrected) group distance.}
\tablenotetext{c}{Mean heliocentric group velocity.}
\tablenotetext{d}{Line-of-sight velocity dispersion.}
\tablenotetext{e}{Projected virial radius.}
\tablenotetext{f}{Log of the virial mass in solar units.}
\tablenotetext{g}{Log of the projected mass in solar units.}
\tablenotetext{i}{Log of the (projected) mass-to-light ratio in solar units.}
\tablenotetext{j}{Composition of group based on known galaxy clusters and superclusters (lists only those groups that appear in HDC catalog with 25 or more members).}
\tablecomments{The LDC group catalog was created using \ParamMax, corresponding to a density contrast \DCMax. We assume \hvalue~where a value is required.}
\end{deluxetable}
\clearpage
\end{landscape}
% Table A4
\clearpage
\begin{landscape}
\begin{deluxetable}{ccccccccccccc}
\tabletypesize{\tiny}
\tablecolumns{13}
%\rotate
\tablewidth{0pt}
\tablecaption{\label{Tab:GroupCatalogBigStd}Groups in the HDC catalog with 25 or more members}
\tablehead{
\colhead{\#} & \colhead{RA} & \colhead{Dec} & \colhead{Members\tablenotemark{a}} & \colhead{Distance\tablenotemark{b}} & \colhead{${V_G}$\tablenotemark{c}} & \colhead{${\sigma_P}$\tablenotemark{d}} & \colhead{${R_\mathrm{P}}$\tablenotemark{e}} & \colhead{$\log \left[ \frac{M_V}{M_\odot} \right]$\tablenotemark{f}} & \colhead{$\log \left[ \frac{M_P}{M_\odot} \right]$\tablenotemark{g}} & \colhead{$\log \left[ \frac{M_P / L}{M_\odot / L_\odot} \right]$\tablenotemark{i}} & \colhead{Corresponding} & \colhead{Identified} \\
 & & & & \colhead{(Mpc)} & \colhead{(km/s)} & \colhead{(km/s)} & \colhead{(Mpc)} & & & & \colhead{Group \#\tablenotemark{j}} & \colhead{With\tablenotemark{k}}
}
\startdata
  716 & 12$^h$33$^m$06.5$^s$ & $+$07$^\circ$47$'$52$''$ &     298 (0) &   19.33 &  1353 &   648.7 &  2.95 &  15.134 &  15.257 &   2.171 &   852 & Virgo \\
  214 & 03$^h$10$^m$24.3$^s$ & $+$41$^\circ$34$'$04$''$ &     172 (0) &   72.21 &  5310 &   974.2 &  2.61 &  15.434 &  15.769 &   2.429 &   229 & Perseus-Pisces (A426) \\
  699 & 12$^h$08$^m$34.1$^s$ & $+$46$^\circ$13$'$02$''$ &     123 (0) &   15.30 &   966 &   405.0 &  3.39 &  14.785 &  15.109 &   2.548 &   852 & Virgo \\
  722 & 12$^h$44$^m$48.4$^s$ & $-$41$^\circ$05$'$22$''$ &     100 (0) &   53.67 &  3452 &   752.1 &  1.95 &  15.083 &  15.193 &   2.097 &   881 & Centaurus (A3526) \\
  928 & 16$^h$15$^m$53.1$^s$ & $-$60$^\circ$54$'$29$''$ &      90 (0) &   69.67 &  4887 &   827.4 &  1.46 &  15.039 &  14.997 &   1.852 &  1117 & Norma (A3627) \\
  592 & 10$^h$36$^m$54.5$^s$ & $-$27$^\circ$11$'$21$''$ &      67 (0) &   51.97 &  3654 &   523.0 &  1.48 &  14.648 &  14.637 &   1.927 &   712 & Hydra (A1060) \\
 1034 & 18$^h$47$^m$40.8$^s$ & $-$63$^\circ$26$'$02$''$ &      51 (0) &   63.29 &  4450 &   350.1 &  1.75 &  14.372 &  14.476 &   1.665 &  1259 &  \\
  237 & 03$^h$38$^m$20.8$^s$ & $-$20$^\circ$34$'$53$''$ &      51 (0) &   21.77 &  1624 &   231.8 &  1.79 &  14.022 &  14.106 &   1.905 &   391 & Eridanus \\
  235 & 03$^h$35$^m$11.8$^s$ & $-$35$^\circ$04$'$15$''$ &      43 (0) &   19.11 &  1438 &   293.2 &  1.02 &  13.981 &  14.094 &   1.721 &   391 & Fornax I \\
  734 & 12$^h$58$^m$51.0$^s$ & $+$27$^\circ$51$'$00$''$ &      42 (0) &  102.23 &  6986 &   628.2 &  1.70 &  14.866 &  14.955 &   1.815 &   890 & Coma (A1656) \\
  669 & 11$^h$44$^m$35.4$^s$ & $+$19$^\circ$58$'$45$''$ &      42 (0) &   95.31 &  6517 &   614.5 &  1.22 &  14.701 &  14.775 &   1.773 &   811 & Coma (A1367) \\
  553 & 10$^h$00$^m$36.8$^s$ & $-$31$^\circ$21$'$56$''$ &      40 (0) &   37.97 &  2706 &   218.4 &  2.67 &  14.144 &  14.371 &   1.853 &   712 & Hydra (A1060) \\
  628 & 11$^h$09$^m$31.5$^s$ & $+$15$^\circ$23$'$42$''$ &      39 (0) &   13.04 &  1035 &   235.2 &  1.16 &  13.845 &  14.082 &   2.164 &   852 & NGC3607 Cluster \\
  103 & 01$^h$53$^m$36.6$^s$ & $+$36$^\circ$17$'$59$''$ &      39 (0) &   66.12 &  4891 &   412.0 &  1.55 &  14.461 &  14.466 &   1.770 &   127 & Perseus-Pisces (A262) \\
  828 & 14$^h$01$^m$45.9$^s$ & $-$33$^\circ$50$'$04$''$ &      36 (0) &   55.71 &  4203 &   367.7 &  1.59 &  14.371 &  14.419 &   1.934 &   985 &  \\
   73 & 01$^h$23$^m$18.6$^s$ & $+$33$^\circ$34$'$59$''$ &      35 (0) &   66.89 &  4953 &   526.0 &  1.01 &  14.486 &  14.429 &   1.766 &    79 &  \\
  584 & 10$^h$30$^m$25.5$^s$ & $-$35$^\circ$20$'$24$''$ &      30 (0) &   40.47 &  2928 &   383.0 &  0.69 &  14.047 &  13.935 &   1.619 &   712 &  \\
  727 & 12$^h$52$^m$38.0$^s$ & $-$08$^\circ$57$'$53$''$ &      29 (0) &   59.59 &  4070 &   350.8 &  1.34 &  14.258 &  14.367 &   1.943 &   883 &  \\
  141 & 02$^h$25$^m$24.5$^s$ & $+$42$^\circ$05$'$38$''$ &      29 (0) &   76.52 &  5642 &   563.1 &  1.36 &  14.675 &  14.801 &   2.170 &   229 & Perseus-Pisces (A347) \\
   78 & 01$^h$26$^m$33.5$^s$ & $-$01$^\circ$34$'$42$''$ &      29 (0) &   70.73 &  5293 &   442.7 &  0.98 &  14.321 &  14.500 &   1.940 &    95 & A194 \\
   56 & 01$^h$09$^m$52.6$^s$ & $+$32$^\circ$43$'$06$''$ &      29 (0) &   68.38 &  5064 &   368.8 &  1.30 &  14.286 &  14.442 &   1.808 &    79 &  \\
  807 & 13$^h$48$^m$47.9$^s$ & $-$30$^\circ$18$'$54$''$ &      28 (0) &   64.59 &  4596 &   432.8 &  1.20 &  14.391 &  14.387 &   1.751 &   956 & Centaurus (A3574) \\
  763 & 13$^h$17$^m$41.1$^s$ & $-$16$^\circ$36$'$08$''$ &      28 (0) &   33.44 &  2332 &   496.4 &  1.09 &  14.469 &  14.900 &   2.759 &   852 &  \\
  432 & 07$^h$21$^m$47.7$^s$ & $-$30$^\circ$04$'$19$''$ &      28 (4) &   28.69 &  2054 &   204.9 &  2.19 &  14.004 &  14.221 &   2.247 &   391 & NGC2280 Cluster \\
  282 & 04$^h$22$^m$27.8$^s$ & $+$36$^\circ$43$'$26$''$ &      28 (0) &   82.20 &  6032 &   266.6 &  1.79 &  14.145 &  14.255 &   1.520 &   292 &  \\
  732 & 12$^h$56$^m$09.5$^s$ & $-$13$^\circ$37$'$44$''$ &      27 (0) &   65.56 &  4510 &   319.2 &  1.34 &  14.175 &  14.529 &   1.926 &   883 &  \\
  295 & 04$^h$35$^m$49.4$^s$ & $-$58$^\circ$57$'$46$''$ &      26 (0) &   15.84 &  1226 &   188.9 &  1.50 &  13.769 &  14.019 &   2.077 &   391 & Dorado \\
\enddata
\tablenotetext{a}{No. of group members (including those generated from the population of the plane). The number derived from the galactic-plane population is contained in parentheses.}
\tablenotetext{b}{Mean (corrected) group distance.}
\tablenotetext{c}{Mean heliocentric group velocity.}
\tablenotetext{d}{Line-of-sight velocity dispersion.}
\tablenotetext{e}{Projected virial radius.}
\tablenotetext{f}{Log of the virial mass in solar units.}
\tablenotetext{g}{Log of the projected mass in solar units.}
\tablenotetext{i}{Log of the (projected) mass-to-light ratio in solar units.}
\tablenotetext{j}{Group number from LDC catalog that encompasses all members of this group.}
\tablenotetext{k}{Composition of group based on known galaxy clusters and superclusters.}
\tablecomments{The HDC group catalog was created using \ParamStd, corresponding to a density contrast \DCStd.
We assume \hvalue~where a value is required.}
\end{deluxetable}
\clearpage
\end{landscape}
% Table A5
\begin{table*}
\begin{center}
\caption{\label{Tab:GalaxiesInGroupsMax}Groups in the LDC catalog and their members}
\begin{tabular}{ccccccc}
\tableline\tableline
Name & RA & Dec & ${V_\mathrm{h}}$\tablenotemark{a} & ${m_K}$\tablenotemark{b} & Distance\tablenotemark{c} & Corresponding \\
 & & & (km/s) & & (Mpc) & Group \#\tablenotemark{d} \\
\tableline
\multicolumn{7}{c}{Group 1} \\
\tableline
000009.14+3244182 & 00$^h$00$^m$09.0$^s$ & $+$32$^\circ$44$'$18$''$ & 10372 &  10.61 &  140.48 &     1 \\
000028.80+3246563 & 00$^h$00$^m$28.8$^s$ & $+$32$^\circ$46$'$56$''$ &  9803 &  11.09 &  132.77 &     1 \\
235950.52+3242086 & 23$^h$59$^m$50.5$^s$ & $+$32$^\circ$42$'$09$''$ & 10086 &  11.12 &  136.61 &     1 \\
\tableline
\multicolumn{7}{c}{Group 2} \\
\tableline
000104.78+0432261 & 00$^h$01$^m$04.7$^s$ & $+$04$^\circ$32$'$26$''$ &  9151 &  11.09 &  123.42 & None  \\
000221.65+0405230 & 00$^h$02$^m$21.6$^s$ & $+$04$^\circ$05$'$23$''$ &  8606 &  11.10 &  116.03 & None  \\
235858.87+0338045 & 23$^h$58$^m$58.8$^s$ & $+$03$^\circ$38$'$04$''$ &  8854 &  10.93 &  119.43 & None  \\
\tableline
\multicolumn{7}{c}{Group 3} \\
\tableline
000037.94+2823041 & 00$^h$00$^m$37.8$^s$ & $+$28$^\circ$23$'$04$''$ &  8705 &  10.46 &  117.75 &  1258 \\
000046.96+2824071 & 00$^h$00$^m$46.8$^s$ & $+$28$^\circ$24$'$07$''$ &  8764 &  10.41 &  118.55 &  1258 \\
000433.73+2818059 & 00$^h$04$^m$33.6$^s$ & $+$28$^\circ$18$'$06$''$ &  8785 &  10.62 &  118.79 & None  \\
235828.41+2802025 & 23$^h$58$^m$28.3$^s$ & $+$28$^\circ$02$'$03$''$ &  9145 &  10.94 &  123.72 &  1258 \\
235943.72+2817251 & 23$^h$59$^m$43.6$^s$ & $+$28$^\circ$17$'$25$''$ &  9073 &  10.71 &  122.74 &  1258 \\
\tableline
\multicolumn{7}{c}{Group 4} \\
\tableline
000100.43+0614312 & 00$^h$01$^m$00.3$^s$ & $+$06$^\circ$14$'$31$''$ &  5324 &  10.03 &   71.77 & None  \\
000348.85+0728429 & 00$^h$03$^m$48.9$^s$ & $+$07$^\circ$28$'$43$''$ &  5241 &   9.67 &   70.63 & None  \\
000649.47+0837425 & 00$^h$06$^m$49.5$^s$ & $+$08$^\circ$37$'$42$''$ &  5257 &  10.58 &   70.82 & None  \\
235651.54+0530303 & 23$^h$56$^m$51.6$^s$ & $+$05$^\circ$30$'$30$''$ &  5405 &  11.05 &   72.90 & None  \\
\tableline
\multicolumn{7}{c}{Group 5} \\
\tableline
000105.97-5359303 & 00$^h$01$^m$06.0$^s$ & $-$53$^\circ$59$'$30$''$ &  9423 &  10.57 &  129.47 & None  \\
000310.64-5444562 & 00$^h$03$^m$10.6$^s$ & $-$54$^\circ$44$'$56$''$ &  9767 &  10.34 &  134.20 & None  \\
000311.27-5444588 & 00$^h$03$^m$11.3$^s$ & $-$54$^\circ$44$'$59$''$ &  9767 &  10.35 &  134.20 & None  \\
\tableline
\multicolumn{7}{c}{Group 6} \\
\tableline
000329.22+2721063 & 00$^h$03$^m$29.1$^s$ & $+$27$^\circ$21$'$06$''$ &  7690 &  11.02 &  103.97 &     4 \\
000548.43+2726579 & 00$^h$05$^m$48.3$^s$ & $+$27$^\circ$26$'$58$''$ &  7531 &  10.95 &  101.81 &     4 \\
000717.10+2740421 & 00$^h$07$^m$17.1$^s$ & $+$27$^\circ$40$'$42$''$ &  7550 &  11.13 &  102.05 &     4 \\
\tableline
\end{tabular}
\end{center}
\tablenotetext{a}{Heliocentric velocity.}
\tablenotetext{b}{Corrected distance, assuming \hvalue.}
\tablenotetext{c}{Apparent K magnitude.}
\tablenotetext{d}{Corresponding group number assigned to this galaxy when in the HDC catalog.}
\tablecomments{The complete version of this table is in the electronic edition of the Journal. The table in the printed edition is only intended as a guide to its content.
The LDC group catalog was created using \ParamMax, corresponding to a density contrast \DCMax.}
\end{table*}

% Table A6
\begin{table*}
\begin{center}
\caption{\label{Tab:GalaxiesInGroupsStd}Groups in the HDC catalog and their members}
\begin{tabular}{ccccccc}
\tableline\tableline
Name & RA & Dec & ${V_\mathrm{h}}$\tablenotemark{a} & ${m_K}$\tablenotemark{b} & Distance\tablenotemark{c} & Corresponding \\
 & & & (km/s) & & (Mpc) & Group \#\tablenotemark{d} \\
\tableline
\multicolumn{7}{c}{Group 1} \\
\tableline
000009.14+3244182 & 00$^h$00$^m$09.0$^s$ & $+$32$^\circ$44$'$18$''$ & 10372 &  10.61 &  140.48 &     1 \\
000028.80+3246563 & 00$^h$00$^m$28.8$^s$ & $+$32$^\circ$46$'$56$''$ &  9803 &  11.09 &  132.77 &     1 \\
235950.52+3242086 & 23$^h$59$^m$50.5$^s$ & $+$32$^\circ$42$'$09$''$ & 10086 &  11.12 &  136.61 &     1 \\
\tableline
\multicolumn{7}{c}{Group 2} \\
\tableline
000001.68+4716282 & 00$^h$00$^m$01.7$^s$ & $+$47$^\circ$16$'$28$''$ &  5017 &   9.68 &   68.91 &    11 \\
000012.95+4657543 & 00$^h$00$^m$13.0$^s$ & $+$46$^\circ$57$'$54$''$ &  5366 &  10.84 &   73.57 &    11 \\
000426.65+4729250 & 00$^h$04$^m$26.6$^s$ & $+$47$^\circ$29$'$25$''$ &  5269 &  10.52 &   72.27 &    11 \\
000527.96+4632371 & 00$^h$05$^m$28.0$^s$ & $+$46$^\circ$32$'$37$''$ &  4971 &  11.02 &   68.21 &    11 \\
000723.79+4702265 & 00$^h$07$^m$23.8$^s$ & $+$47$^\circ$02$'$27$''$ &  5313 &   9.93 &   72.81 &    11 \\
000724.58+4659195 & 00$^h$07$^m$24.6$^s$ & $+$46$^\circ$59$'$20$''$ &  5097 &  10.87 &   69.91 &    11 \\
235247.40+4648138 & 23$^h$52$^m$47.3$^s$ & $+$46$^\circ$48$'$14$''$ &  5047 &  10.65 &   69.35 &    11 \\
235401.14+4729225 & 23$^h$54$^m$01.1$^s$ & $+$47$^\circ$29$'$22$''$ &  5202 &  10.39 &   71.45 &    11 \\
235526.14+4716485 & 23$^h$55$^m$26.1$^s$ & $+$47$^\circ$16$'$49$''$ &  5348 &  10.98 &   73.38 &    11 \\
235915.79+4653213 & 23$^h$59$^m$15.8$^s$ & $+$46$^\circ$53$'$21$''$ &  5021 &   9.46 &   68.95 &    11 \\
\tableline
\multicolumn{7}{c}{Group 3} \\
\tableline
000126.77+3126016 & 00$^h$01$^m$26.7$^s$ & $+$31$^\circ$26$'$02$''$ &  4948 &  10.23 &   67.25 &     9 \\
000130.05+3126306 & 00$^h$01$^m$30.0$^s$ & $+$31$^\circ$26$'$31$''$ &  4767 &  10.47 &   64.83 &     9 \\
000308.87+3102108 & 00$^h$03$^m$08.9$^s$ & $+$31$^\circ$02$'$11$''$ &  4797 &  11.16 &   65.20 &     9 \\
000424.49+3128193 & 00$^h$04$^m$24.5$^s$ & $+$31$^\circ$28$'$19$''$ &  4958 &  11.14 &   67.36 &     9 \\
\tableline
\multicolumn{7}{c}{Group 4} \\
\tableline
000329.22+2721063 & 00$^h$03$^m$29.1$^s$ & $+$27$^\circ$21$'$06$''$ &  7690 &  11.02 &  103.97 &     6 \\
000548.43+2726579 & 00$^h$05$^m$48.3$^s$ & $+$27$^\circ$26$'$58$''$ &  7531 &  10.95 &  101.81 &     6 \\
000717.10+2740421 & 00$^h$07$^m$17.1$^s$ & $+$27$^\circ$40$'$42$''$ &  7550 &  11.13 &  102.05 &     6 \\
\tableline
\multicolumn{7}{c}{Group 5} \\
\tableline
000457.78+0507245 & 00$^h$04$^m$57.8$^s$ & $+$05$^\circ$07$'$24$''$ &  5357 &  11.19 &   72.17 &     7 \\
000527.66+0513204 & 00$^h$05$^m$27.6$^s$ & $+$05$^\circ$13$'$20$''$ &  5294 &  10.08 &   71.31 &     7 \\
000640.35+0506483 & 00$^h$06$^m$40.3$^s$ & $+$05$^\circ$06$'$48$''$ &  5371 &  11.22 &   72.34 &     7 \\
\tableline
\end{tabular}
\end{center}
\tablenotetext{a}{Heliocentric velocity.}
\tablenotetext{b}{Corrected distance, assuming \hvalue.}
\tablenotetext{c}{Apparent K magnitude.}
\tablenotetext{d}{Corresponding group number assigned to this galaxy in the LDC catalog.}
\tablecomments{The complete version of this table is in the electronic edition of the Journal. The table in the printed edition is only intended as a guide to its content.
The HDC group catalog was created using \ParamStd, corresponding to a density contrast \DCStd.}
\end{table*}

%\clearpage

\clearpage

%\bibliographystyle{apj}
%\bibliography{references}

\begin{thebibliography}{}

\bibitem[{{Bahcall} \& {Tremaine}(1981){Bahcall} \& {Tremaine}}]{Bahcall:1981}
{Bahcall}, J.~N., \& {Tremaine}, S. 1981, \apj, 244, 805

\bibitem[{{Bell} {et~al.}(2003){Bell}, {McIntosh}, {Katz}, \&  {Weinberg}}]{Bell:2003}
{Bell}, E.~F., {McIntosh}, D.~H., {Katz}, N., \& {Weinberg}, M.~D. 2003, \apjs,  149, 289

\bibitem[{{Bennett} {et~al.}(1996){Bennett}, {Banday}, {Gorski}, {Hinshaw},  {Jackson}, {Keegstra}, {Kogut}, {Smoot}, {Wilkinson}, \&  {Wright}}]{Bennett:1996}
{Bennett}, C.~L. {et~al.} 1996, \apjl, 464, L1

\bibitem[{{Bennett} {et~al.}(2003){Bennett}, {Halpern}, {Hinshaw}, {Jarosik},  {Kogut}, {Limon}, {Meyer}, {Page}, {Spergel}, {Tucker}, {Wollack}, {Wright},  {Barnes}, {Greason}, {Hill}, {Komatsu}, {Nolta}, {Odegard}, {Peiris},  {Verde}, \& {Weiland}}]{Bennett:2003}
---. 2003, \apjs, 148, 1

\bibitem[{{Branchini} {et~al.}(1999){Branchini}, {Teodoro}, {Frenk},  {Schmoldt}, {Efstathiou}, {White}, {Saunders}, {Sutherland},  {Rowan-Robinson}, {Keeble}, {Tadros}, {Maddox}, \& {Oliver}}]{Branchini:1999}
{Branchini}, E. {et~al.} 1999, \mnras, 308, 1

\bibitem[{{Chakravarti} {et~al.}(1967){Chakravarti}, {Laha}, \&  {Roy}}]{Chakravarti:1967}
{Chakravarti}, {Laha}, \& {Roy}. 1967, {Handbook of Methods of Applied  Statistics}, Vol.~I (John Wiley and Sons), 392--394

\bibitem[{{Courteau} \& {van den Bergh}(1999){Courteau} \& {van den Bergh}}]{Courteau:1999}
{Courteau}, S., \& {van den Bergh}, S. 1999, \aj, 118, 337

\bibitem[{{Cutri} {et~al.}(2003){Cutri}}]{Cutri:2003}
{Cutri}, R., {et~al.} 2003, IPAC, http://www.ipac.caltech.edu/2mass/

\bibitem[{{da Costa} {et~al.}(1998){da Costa}, {Willmer}, {Pellegrini},  {Chaves}, {Rit{\'e}}, {Maia}, {Geller}, {Latham}, {Kurtz}, {Huchra},  {Ramella}, {Fairall}, {Smith}, \& {L{\'{\i}}pari}}]{daCosta:1998}
{da Costa}, L.~N. {et~al.} 1998, \aj, 116, 1

\bibitem[{{Davis} {et~al.}(2003){Davis}, {Faber}, {Newman}, {Phillips},  {Ellis}, {Steidel}, {Conselice}, {Coil}, {Finkbeiner}, {Koo}, {Guhathakurta},  {Weiner}, {Schiavon}, {Willmer}, {Kaiser}, {Luppino}, {Wirth}, {Connolly},  {Eisenhardt}, {Cooper}, \& {Gerke}}]{Davis:2003}
{Davis}, M. {et~al.} 2003, in Discoveries and Research Prospects from 6- to  10-Meter-Class Telescopes II. Edited by Guhathakurta, Puragra. Proceedings of  the SPIE, Volume 4834, pp. 161-172 (2003)., ed. P.~{Guhathakurta}, 161--172

\bibitem[{{de Lapparent} {et~al.}(1991){de Lapparent}, {Geller}, \&  {Huchra}}]{deLapparent:1991}
{de Lapparent}, V., {Geller}, M.~J., \& {Huchra}, J.~P. 1991, \apj, 369, 273

\bibitem[{{de Vaucouleurs}(1975){de Vaucouleurs}}]{deVaucouleurs:1975}
{de Vaucouleurs}, G. 1975, Stars and stellar systems, 9, 557

\bibitem[{{Diaferio} {et~al.}(1999){Diaferio}, {Kauffmann}, {Colberg}, \&  {White}}]{Diaferio:1999}
{Diaferio}, A., {Kauffmann}, G., {Colberg}, J.~M., \& {White}, S.~D.~M. 1999,  \mnras, 307, 537

\bibitem[{{Eke} {et~al.}(2004){Eke}, {Baugh}, {Cole}, {Frenk}, {Norberg},  {Peacock}, {Baldry}, {Bland-Hawthorn}, {Bridges}, {Cannon}, {Colless},  {Collins}, {Couch}, {Dalton}, {de Propris}, {Driver}, {Efstathiou}, {Ellis},  {Glazebrook}, {Jackson}, {Lahav}, {Lewis}, {Lumsden}, {Maddox}, {Madgwick},  {Peterson}, {Sutherland}, \& {Taylor}}]{Eke:2004}
{Eke}, V.~R. {et~al.} 2004, \mnras, 348, 866

\bibitem[{{Erdo{\u g}du} {et~al.}(2006){Erdo{\u g}du}, {Lahav}, {Huchra},  {Colless}, {Cutri}, {Falco}, {George}, {Jarrett}, {Jones}, {Macri}, {Mader},  {Martimbeau}, {Pahre}, {Parker}, {Rassat}, \& {Saunders}}]{Erdogdu:2006b}
{Erdo{\u g}du}, P. {et~al.} 2006, \mnras

\bibitem[{{Falco} {et~al.}(1999){Falco}, {Kurtz}, {Geller}, {Huchra}, {Peters},  {Berlind}, {Mink}, {Tokarz}, \& {Elwell}}]{Falco:1999}
{Falco}, E.~E. {et~al.} 1999, VizieR Online Data Catalog, 611, 10438

\bibitem[{{Fisher} {et~al.}(1995){Fisher}, {Huchra}, {Strauss}, {Davis},  {Yahil}, \& {Schlegel}}]{Fisher:1995}
{Fisher}, K.~B., {Huchra}, J.~P., {Strauss}, M.~A., {Davis}, M., {Yahil}, A.,  \& {Schlegel}, D. 1995, \apjs, 100, 69

\bibitem[{{Frederic}(1995a){Frederic}}]{Frederic:1995b}
{Frederic}, J.~J. 1995a, \apjs, 97, 275

\bibitem[{{Frederic}(1995b){Frederic}}]{Frederic:1995a}
---. 1995b, \apjs, 97, 259

\bibitem[{{Geller} \& {Huchra}(1983){Geller} \& {Huchra}}]{Geller:1983}
{Geller}, M.~J., \& {Huchra}, J.~P. 1983, \apjs, 52, 61

\bibitem[{{Geller} \& {Huchra}(1989){Geller} \& {Huchra}}]{Geller:1989}
---. 1989, Science, 246, 897

\bibitem[{{Gerke} {et~al.}(2005){Gerke}, {Newman}, {Davis}, {Marinoni}, {Yan},  {Coil}, {Conroy}, {Cooper}, {Faber}, {Finkbeiner}, {Guhathakurta}, {Kaiser},  {Koo}, {Phillips}, {Weiner}, \& {Willmer}}]{Gerke:2005}
{Gerke}, B.~F. {et~al.} 2005, \apj, 625, 6

\bibitem[{{Giuricin} {et~al.}(2000){Giuricin}, {Marinoni}, {Ceriani}, \&  {Pisani}}]{Giuricin:2000}
{Giuricin}, G., {Marinoni}, C., {Ceriani}, L., \& {Pisani}, A. 2000, \apj, 543,  178

\bibitem[{{Heisler} {et~al.}(1985){Heisler}, {Tremaine}, \&  {Bahcall}}]{Heisler:1985}
{Heisler}, J., {Tremaine}, S., \& {Bahcall}, J.~N. 1985, \apj, 298, 8

\bibitem[{{Huchra} {et~al.}(2005a){Huchra}, {Jarrett},  {Skrutskie}, {Cutri}, {Schneider}, {Macri}, {Steining}, {Mader},  {Martimbeau}, \& {George}}]{Huchra:2005b}
{Huchra}, J. {et~al.} 2005a, in ASP Conf. Ser. 329: Nearby  Large-Scale Structures and the Zone of Avoidance, ed. A.~P. {Fairall} \&  P.~A. {Woudt}, 135

\bibitem[{{Huchra} {et~al.}(2005b){Huchra}, {Martimbeau},  {Jarrett}, {Cutri}, {Skrutskie}, {Schneider}, {Steining}, {Macri}, {Mader},  \& {George}}]{Huchra:2005}
{Huchra}, J. {et~al.} 2005b, in IAU Symposium No. 216, 170

\bibitem[{{Huchra} \& {Geller}(1982){Huchra} \& {Geller}}]{Huchra:1982}
{Huchra}, J.~P., \& {Geller}, M.~J. 1982, \apj, 257, 423

\bibitem[{{Huchra} {et~al.}(1995){Huchra}, {Geller}, \& {Corwin}}]{Huchra:1995}
{Huchra}, J.~P., {Geller}, M.~J., \& {Corwin}, H.~G. 1995, \apjs, 99, 391

\bibitem[{{Huchra} {et~al.}(1990){Huchra}, {Geller}, {de Lapparent}, \&  {Corwin}}]{Huchra:1990}
{Huchra}, J.~P., {Geller}, M.~J., {de Lapparent}, V., \& {Corwin}, H.~G. 1990,  \apjs, 72, 433

\bibitem[{{Jarrett} {et~al.}(2000){Jarrett}, {Chester}, {Cutri}, {Schneider},  {Skrutskie}, \& {Huchra}}]{Jarrett:2000}
{Jarrett}, T.~H., {Chester}, T., {Cutri}, R., {Schneider}, S., {Skrutskie}, M.,  \& {Huchra}, J.~P. 2000, \aj, 119, 2498

\bibitem[{{Jenkins} {et~al.}(2001){Jenkins}, {Frenk}, {White}, {Colberg},  {Cole}, {Evrard}, {Couchman}, \& {Yoshida}}]{Jenkins:2001}
{Jenkins}, A., {Frenk}, C.~S., {White}, S.~D.~M., {Colberg}, J.~M., {Cole}, S.,  {Evrard}, A.~E., {Couchman}, H.~M.~P., \& {Yoshida}, N. 2001, \mnras, 321,  372

\bibitem[{{Johnson}(1966){Johnson}}]{Johnson:1966}
{Johnson}, H.~L. 1966, \araa, 4, 193

\bibitem[{{Kochanek} {et~al.}(2001){Kochanek}, {Pahre}, {Falco}, {Huchra},  {Mader}, {Jarrett}, {Chester}, {Cutri}, \& {Schneider}}]{Kochanek:2001}
{Kochanek}, C.~S. {et~al.} 2001, \apj, 560, 566

\bibitem[{{Kochanek} {et~al.}(2003){Kochanek}, {White}, {Huchra}, {Macri},  {Jarrett}, {Schneider}, \& {Mader}}]{Kochanek:2003}
{Kochanek}, C.~S., {White}, M., {Huchra}, J., {Macri}, L., {Jarrett}, T.~H.,  {Schneider}, S.~E., \& {Mader}, J. 2003, \apj, 585, 161

\bibitem[{{Lahav} {et~al.}(1994){Lahav}, {Fisher}, {Hoffman}, {Scharf}, \&  {Zaroubi}}]{Lahav:1994}
{Lahav}, O., {Fisher}, K.~B., {Hoffman}, Y., {Scharf}, C.~A., \& {Zaroubi}, S.  1994, \apjl, 423, L93

\bibitem[{{Marinoni} {et~al.}(2002){Marinoni}, {Davis}, {Newman}, \&  {Coil}}]{Marinoni:2002}
{Marinoni}, C., {Davis}, M., {Newman}, J.~A., \& {Coil}, A.~L. 2002, \apj, 580,  122

\bibitem[{{Mart{\'{\i}}nez} {et~al.}(2002){Mart{\'{\i}}nez}, {Zandivarez},  {Merch{\'a}n}, \& {Dom{\'{\i}}nguez}}]{Martinez:2002}
{Mart{\'{\i}}nez}, H.~J., {Zandivarez}, A., {Merch{\'a}n}, M.~E., \&  {Dom{\'{\i}}nguez}, M.~J.~L. 2002, \mnras, 337, 1441

\bibitem[{{Merch{\'a}n} \& {Zandivarez}(2002){Merch{\'a}n} \& {Zandivarez}}]{Merchan:2002}
{Merch{\'a}n}, M., \& {Zandivarez}, A. 2002, \mnras, 335, 216

\bibitem[{{Merch{\'a}n} \& {Zandivarez}(2005){Merch{\'a}n} \& {Zandivarez}}]{Merchan:2005}
{Merch{\'a}n}, M.~E., \& {Zandivarez}, A. 2005, \apj, 630, 759

\bibitem[{{Miller} {et~al.}(2005){Miller}, {Nichol}, {Reichart}, {Wechsler},  {Evrard}, {Annis}, {McKay}, {Bahcall}, {Bernardi}, {Boehringer}, {Connolly},  {Goto}, {Kniazev}, {Lamb}, {Postman}, {Schneider}, {Sheth}, \&  {Voges}}]{Miller:2005}
{Miller}, C.~J. {et~al.} 2005, \aj, 130, 968

\bibitem[{{Mould} {et~al.}(2000){Mould}, {Huchra}, {Freedman}, {Kennicutt},  {Ferrarese}, {Ford}, {Gibson}, {Graham}, {Hughes}, {Illingworth}, {Kelson},  {Macri}, {Madore}, {Sakai}, {Sebo}, {Silbermann}, \& {Stetson}}]{Mould:2000}
{Mould}, J.~R. {et~al.} 2000, \apj, 529, 786

\bibitem[{{Nolthenius} \& {White}(1987){Nolthenius} \& {White}}]{Nolthenius:1987}
{Nolthenius}, R., \& {White}, S.~D.~M. 1987, \mnras, 225, 505

\bibitem[{{Press} \& {Schechter}(1974){Press} \& {Schechter}}]{Press:1974}
{Press}, W.~H., \& {Schechter}, P. 1974, \apj, 187, 425

\bibitem[{{Ramella} {et~al.}(1989){Ramella}, {Geller}, \&  {Huchra}}]{Ramella:1989}
{Ramella}, M., {Geller}, M.~J., \& {Huchra}, J.~P. 1989, \apj, 344, 57

\bibitem[{{Ramella} {et~al.}(2002){Ramella}, {Geller}, {Pisani}, \& {da  Costa}}]{Ramella:2002}
{Ramella}, M., {Geller}, M.~J., {Pisani}, A., \& {da Costa}, L.~N. 2002, \aj,  123, 2976

\bibitem[{{Ramella} {et~al.}(1997){Ramella}, {Pisani}, \&  {Geller}}]{Ramella:1997}
{Ramella}, M., {Pisani}, A., \& {Geller}, M.~J. 1997, \aj, 113, 483

\bibitem[{{Saunders} {et~al.}(2000){Saunders}, {Sutherland}, {Maddox},  {Keeble}, {Oliver}, {Rowan-Robinson}, {McMahon}, {Efstathiou}, {Tadros},  {White}, {Frenk}, {Carrami{\~n}ana}, \& {Hawkins}}]{Saunders:2000}
{Saunders}, W. {et~al.} 2000, \mnras, 317, 55

\bibitem[{{Schechter}(1976){Schechter}}]{Schechter:1976}
{Schechter}, P. 1976, \apj, 203, 297

\bibitem[{{Schlegel} {et~al.}(1998){Schlegel}, {Finkbeiner}, \&  {Davis}}]{Schlegel:1998}
{Schlegel}, D.~J., {Finkbeiner}, D.~P., \& {Davis}, M. 1998, \apj, 500, 525

\bibitem[{{Sheth} \& {Tormen}(1999){Sheth} \& {Tormen}}]{Sheth:1999}
{Sheth}, R.~K., \& {Tormen}, G. 1999, \mnras, 308, 119

\bibitem[{{Skrutskie} {et~al.}(2006){Skrutskie}, {Cutri}, {Stiening},  {Weinberg}, {Schneider}, {Carpenter}, {Beichman}, {Capps}, {Chester},  {Elias}, {Huchra}, {Liebert}, {Lonsdale}, {Monet}, {Price}, {Seitzer},  {Jarrett}, {Kirkpatrick}, {Gizis}, {Howard}, {Evans}, {Fowler}, {Fullmer},  {Hurt}, {Light}, {Kopan}, {Marsh}, {McCallon}, {Tam}, {Van Dyk}, \&  {Wheelock}}]{Skrutskie:2006}
{Skrutskie}, M.~F. {et~al.} 2006, \aj, 131, 1163

\bibitem[{{Spergel} {et~al.}(2006){Spergel}, {Bean}, {Dore'}, {Nolta},  {Bennett}, {Hinshaw}, {Jarosik}, {Komatsu}, {Page}, {Peiris}, {Verde},  {Barnes}, {Halpern}, {Hill}, {Kogut}, {Limon}, {Meyer}, {Odegard}, {Tucker},  {Weiland}, {Wollack}, \& {Wright}}]{Spergel:2006}
{Spergel}, D.~N. {et~al.} 2006, \apj

\bibitem[{{Strauss} {et~al.}(1992){Strauss}, {Davis}, {Yahil}, \&  {Huchra}}]{Strauss:1992a}
{Strauss}, M.~A., {Davis}, M., {Yahil}, A., \& {Huchra}, J.~P. 1992, \apj, 385,  421

\bibitem[{{Turner} \& {Gott}(1976){Turner} \& {Gott}}]{Turner:1976}
{Turner}, E.~L., \& {Gott}, J.~R. 1976, \apjs, 32, 409

\bibitem[{{Weinmann} {et~al.}(2006){Weinmann}, {van den Bosch}, {Yang}, \&  {Mo}}]{Weinmann:2006}
{Weinmann}, S.~M., {van den Bosch}, F.~C., {Yang}, X., \& {Mo}, H.~J. 2006,  \mnras, 366, 2

\bibitem[{{Yahil} {et~al.}(1991){Yahil}, {Strauss}, {Davis}, \&  {Huchra}}]{Yahil:1991}
{Yahil}, A., {Strauss}, M.~A., {Davis}, M., \& {Huchra}, J.~P. 1991, \apj, 372,  380

\bibitem[{{Yahil} {et~al.}(1977){Yahil}, {Tammann}, \& {Sandage}}]{Yahil:1977}
{Yahil}, A., {Tammann}, G.~A., \& {Sandage}, A. 1977, \apj, 217, 903

\bibitem[{{Yang} {et~al.}(2005){Yang}, {Mo}, {van den Bosch}, \&  {Jing}}]{Yang:2005}
{Yang}, X., {Mo}, H.~J., {van den Bosch}, F.~C., \& {Jing}, Y.~P. 2005, \mnras,  356, 1293

\end{thebibliography}

\vspace{10mm}
Due to size limitations imposed by arxiv.org, print-quality figures appear only in the full-resolution PS and PDF versions, which are available at \textsf{http://www.cfa.harvard.edu/{$\sim$}acrook/preprints/}.

\end{document}